\documentclass[12pt,thmsa]{article}
\usepackage{amssymb}

\usepackage{sw20lart}

\input{tcilatex}
\begin{document}

\begin{center}
{\LARGE Solving the quantum search problem in polynomial time on an NMR
quantum computer} 
\[
\]

Xijia Miao

Corresponding address: Center for Magnetic Resonance Research, University of
Minnesota, 2021 Sixth Street SE, Minneapolis, MN 55455, USA; E-mail:
miao@cmrr.umn.edu\\[0pt]
\[
\]

\textbf{Abstract}
\end{center}

The quantum search problem is an important problem due to the fact that a
general NP problem can be solved by an unsorted quantum search algorithm.
Here it has been shown that the quantum search problem could be solved in
polynomial time on an NMR quantum computer. The NMR ensemble quantum
computation is based on the quantum mechanical unitary dynamics that both a
closed quantum system and its ensemble obey the same quantum mechanical
unitary dynamics instead of on the pseudopure state or the effective pure
state of the classical NMR quantum computation. Based on the new principle
the conventional NMR multiple-quantum spectroscopy has been developed to
solve the search problem. The solution information of the search problem is
first loaded on the unitary evolution propagator which is constructed with
the oracle unitary operation and oracle-independent unitary operations and
used to excite the multiple-quantum coherence in a spin ensemble. Then the
multiple-quantum spectroscopy is used to extract experimentally the solution
information. It has been discussed how to enhance the output NMR signal of
the quantum search NMR multiple-quantum pulse  sequence and some approaches
to enhancing the NMR signal are also proposed. The present work could be
helpful for the conventional high magnetic field NMR machines to solve
efficiently the quantum search problem. 
\[
\]
\newline
{\Large 1. Introduction}

An efficient quantum search algorithm may be tremendously useful because of
the wide range of its practical applications, and especially due to the fact
that a general NP problem can be solved by the unsorted quantum search
algorithm [1]. It is believed extensively that a general NP problem can not
be solved efficiently on a classical digital computer in polynomial time.
But if there would be a polynomial-time quantum search algorithm, then a
general NP problem would be efficiently solved by the algorithm in
polynomial time on a quantum computer. In the past years a great effort has
been devoted to searching for an efficient quantum search algorithm. On the
basis of the quantum parallelism principle [2] Grover first proposed a
quantum search algorithm with quadratic speedup over the classical
counterparts [3, 4]. It has been shown that this algorithm is optimal [5,
6]. A large number of works of the Grover algorithm and its applications
have been published in the past several years. However, the quantum search
algorithms [3, 4] which are based on the pure quantum states are not
polynomial-time algorithms and hence have not been proved certainly that
they can move the frontier between solvability and intractability of
complexity problems [6]. On the other hand, the experimental realization of
quantum computation and the physical architecture of quantum computers have
achieved a great progress in the recent years, although it is the long term
to build up a commercial quantum computer. In particular, the experimental
ensemble quantum computation based on the nuclear magnetic resonance (NMR)
techniques [7, 8] has been extensively studied [9, 10] and a large number of
the related works have been reported [11-14]. But the classical NMR quantum
computation [9-15] working on the concept of the pseudopure state [9] or the
effective pure state [10] suffers from many trouble problems [15, 16]. For
example, people suspect whether or not the classical NMR quantum computation
is real quantum computation and is as powerful as the quantum computation
based on pure quantum states [16]. Since the effective pure states or the
pseudopure states are isomorphic to the pure quantum states [9, 10, 12] the
classical NMR quantum computation is at most as powerful as the original one
based on pure quantum states and thus, can not be able to move the frontier
between solvability and intractability. It has to give up many important
characteristic features of a spin ensemble in order that with the help of
the concept of the effective pure state or the pseudopure state the
pure-state the pure-state quantum computation can work in the spin ensemble
as well. For example, highly mixed states of a spin ensemble are not
suitable for the classical NMR quantum computation. It is known that the
quantum states of the spin system in a spin ensemble are very highly mixed.
Therefore, the classical NMR quantum computation can not give a satisfied
answer whether or not the highly mixed states in a spin ensemble could be
used in quantum computation and about the power question of quantum
computation with the highly mixed states.

Amazingly, recently it has been shown that an NMR quantum computer operating
on the highly mixed states of a spin quantum ensemble can be exploited to
solve efficiently the unsorted search problem [17] and the parity problem
[18] in polynomial time, while these problems have been shown to be hard NP
problems and can not be solved efficiently in classical computation and even
in pure-state quantum computation [19, 20]. These results show powerfully
that (a) a general NP problem could be solved in polynomial time on a
quantum computer; (b) quantum entanglement of a quantum system may not be
the sole origin of quantum computation power and the quantum mechanical
unitary dynamics may play an important role in solving these NP problems in
polynomial time; (c) quantum computers can outperform classical computers
and ensemble quantum computation can be at least as powerful as the standard
quantum computation based on pure quantum states. These quantum algorithms
or sequences that operate on highly mixed states in a spin quantum ensemble
could achieve an exponential speedup over the classical search computation
and therefore they really move the frontier between the solvability and
intractability of complex problems. Although in theory the NMR ensemble
quantum computation on highly mixed states is successful to solve the
unsorted quantum search problem and the parity problem in polynomial time,
it still suffers from some problems and limits in practice. The most severe
problem is that the output NMR signal intensity of these sequences reduces
exponentially as the qubit number of the spin system in the spin ensemble
[17, 18], making the NMR ensemble quantum computation available only in spin
systems with a small number of qubits. There may be two schemes to extend
the scale of the NMR quantum computation so that it could solve the search
problem with a larger qubit number. One is to enhance the initial spin
polarization of a spin ensemble by a variety of possible spin-polarization
enhancement techniques [7, 8, 21, 22, 23]. Another is to design new quantum
search algorithms or sequences which consist of a polynomial number of the
oracle unitary operations and oracle-independent unitary operations since
the output NMR signal could be increased by many calls of the oracle unitary
operations in the sequences. In this paper according to the second scheme
new quantum search sequences on the NMR ensemble quantum computers are
proposed to solve the search problem with the help of the multiple-quantum
operator algebra space formalism [24, 25], and particularly the NMR
multiple-quantum spectroscopy [7, 8] has been developed to search for the
unknown marked state, i.e., the solution of the search problem.

There may be two possible schemes for quantum computation making a
transition from a pure-state quantum system to its quantum ensemble. One is
the classical NMR quantum computation. It is based on the concept of the
effective pure state [10] or the pseudopure state [9] which is isomorphic to
the pure quantum state. Then, the quantum computation on the pure state
quantum systems allows to apply to their quantum ensembles with the help of
the concept of the effective pure state. In the frame of the classical NMR
quantum computation information of quantum computation is mainly carried by
the effective pure states. Another is the new NMR ensemble quantum
computation. It is based on the fact that both the closed pure-state quantum
system and its quantum ensemble obey the same quantum mechanical unitary
dynamics when decoherence effects in the quantum system and its ensemble are
negligible [17, 18]. Then in the frame of the NMR ensemble quantum
computation the information of quantum computation is usually loaded on the
unitary evolution propagator that describes the quantum mechanical unitary
dynamics of the quantum system and its ensemble. The highly mixed states
could be used in the ensemble quantum computation because the new ensemble
quantum computation allows the independence each other between the initial
input state and the quantum circuit in a quantum algorithm or sequence.
Therefore, the quantum search process based on the new ensemble quantum
computation is that the information of the unknown marked state of the
search problem is first loaded on the quantum mechanical unitary evolution
propagator, then one uses various experiment and measurement techniques such
as the NMR multiple-quantum spectroscopy to extract the solution information
of the search problem. 
\[
\]
\newline
{\Large 2. The simple quantum search sequences on a spin ensemble}

First of all, the explicit form of the oracle unitary operation of the
quantum search problem is introduced below. In a quantum search algorithm
the basic oracle unitary operation $U_{f}$ is defined as [26]

$\qquad \qquad \qquad U_{f}|x\rangle |a\rangle =|x\rangle |a\bigoplus
f(x)\rangle \qquad \qquad \qquad \qquad \qquad \qquad \quad \ \ (1)$ \newline
where the quantum state $|x\rangle |a\rangle $ is an arbitrary computational
basis of the whole quantum system which consists of the work qubits $I$ and
the auxiliary qubits $S$, the quantum states $|x\rangle $ and $|a\rangle $
are the computational bases of the work qubits $I$ and the auxiliary qubits $%
S$, respectively, and the direction sum symbol $\bigoplus $ denotes addition
modulo $2$. The function $f(x)$ in the quantum search algorithm is defined
over the range $0\leq x\leq N-1$ $(N=2^{n})$ by $f(x)=1$ if $x$ is a
solution to the search problem $(x=s)$, which corresponds to the marked
quantum state $|s\rangle $ of the work qubits $I$, and $f(x)=0$ if $x$ is
not a solution to the search problem $(x\neq s)$.

When the basic oracle unitary operation $U_{f}$ is applied to the quantum
state $|x\rangle |S\rangle $ where the quantum state $|S\rangle $ of the
auxiliary qubits $S$ is chosen as the superposition: $|S\rangle =\frac{1}{%
\sqrt{2}}(|0\rangle -|1\rangle )$ it follows from the definition (1) of the
oracle unitary operation $U_{f}$ that [26]

$\qquad U_{f}|x\rangle |S\rangle =\exp (-i\pi f(x))|x\rangle |S\rangle =\{ 
\begin{array}{l}
-|x\rangle |S\rangle \quad \text{if }x=s \\ 
\ \ |x\rangle |S\rangle \quad \text{if }x\neq s
\end{array}
.\qquad \qquad (2)$\newline
This shows that only the marked quantum state $|s\rangle $ is inverted in
phase or changes sign of its amplitude but any other state $|x\rangle $
keeps unchanged when the function values $f(x)$ are simultaneously evaluated
once over all $x$ values, that is, the oracle unitary operation $U_{f}$ is
performed once on the quantum system. One can see from Eqs.(1) and (2) that
the oracle unitary operation $U_{f}$ is independent of any quantum state $%
|x\rangle $ of the work qubits $I$ but related to the quantum state $%
|S\rangle $ of the auxiliary qubits $S.$ It must be paid much attention to
this point [17, 18] when a quantum search algorithm constructed with the
oracle unitary operation $U_{f}$ and oracle-independent unitary operations
is extended to a quantum spin ensemble from its pure-state quantum spin
system. For convenient design of a quantum search algorithm or sequence, an
auxiliary oracle unitary operation $U_{o}$ is introduced below. This oracle
unitary operation is expressed as $U_{o}(\theta )=U_{f}^{-1}V_{S}(\theta
)U_{f}$ [17] whose unitary operation on any quantum state of the quantum
system is defined by

$U_{o}(\theta )|x\rangle |0\rangle |1\rangle =$ $U_{f}^{-1}V_{S}(\theta
)|x\rangle |0\bigoplus f(x)\rangle |1\rangle $

$=\exp (-i\theta \delta (f(x),1))|x\rangle |0\rangle |1\rangle \qquad \qquad
\qquad \qquad \qquad \qquad \qquad \qquad \quad \ \ (3)$ \newline
where the conditional phase-shift unitary operation $V_{S}(\theta )$ [27]
acts only on the auxiliary quantum state $|S\rangle =|a\rangle |b\rangle $
and is defined by

$V_{S}(\theta )|a\rangle |1\rangle =\exp (-i\theta \delta (a,1))|a\rangle
|1\rangle =\{ 
\begin{array}{l}
\exp (-i\theta )|a\rangle |1\rangle \quad \text{if }a=1 \\ 
\quad \quad \ |a\rangle |1\rangle \qquad \ \ \text{if }a\neq 1
\end{array}
.\quad \ \ (4)$ \newline
Note that the basic oracle unitary operation $U_{f}$ has the property: $%
U_{f}^{2}=E$ (unity operation), as can be seen from Eq.(1). Then the oracle
unitary operation $U_{o}$ can also be expressed as $U_{o}(\theta
)=U_{f}V_{S}(\theta )U_{f}.$

The quantum parallelism principle [2] provides the possibility for quantum
computation outperforming the classical computation. By the quantum
parallelism all the function values $f(x)$ can be evaluated simultaneously
over all the input computational base $|x\rangle $ of the quantum system
when the oracle unitary operation $U_{o}$ acts once on any superposition $%
|\Psi \rangle =\stackrel{N-1}{\stackunder{x=0}{\sum }}a_{x}|x\rangle
|S\rangle $ of the quantum system,

$U_{o}(\theta )|\Psi \rangle =\stackrel{N-1}{\stackunder{x=0}{\sum }}\exp
(-i\theta \delta (f(x),1))a_{x}|x\rangle |S\rangle $

$=\exp (-i\theta )a_{s}|s\rangle |S\rangle +\stackrel{N-1}{\stackunder{%
x=0,x\neq s}{\sum }}a_{x}|x\rangle |S\rangle \ \qquad \qquad \qquad \qquad
\qquad \qquad \qquad (5)$ \newline
where the definition of the function $f(x)$ is used: $f(x)=1$ if $x=s$; $%
f(x)=0$ if $x\neq s$. To analyze conveniently the unitary evolution process
of the quantum system and its ensemble during the action of the oracle
unitary operation $U_{o}$ one needs to use the selective phase-shift
operation $C_{s}(\theta )$ [17] to express explicitly the oracle unitary
operation. The selective phase-shift operation $C_{r}(\theta )$ is a
diagonal unitary operator which is only applied to the work qubits $I$. It
can be expressed in the exponential form

$\qquad \qquad \qquad C_{r}(\theta )=\exp (-i\theta D_{r})\qquad \qquad
\qquad \qquad \qquad \qquad \quad \ \ \qquad (6)$ \newline
where the diagonal operator $D_{r}$ is related to the quantum state $%
|r\rangle $ of the quantum system and is a $LOMSO$ operator [24], and it is
defined as $D_{r}=diag(0,...,0,1,0,...0)$, that is, the diagonal element $%
(D_{r})_{rr}=1$ only for the index $r$ and $(D_{r})_{tt}=0$ for any other
index $t\neq r.$ It has been proved [17] that when the selective phase-shift
operation $C_{s}(\theta )$ of the marked state $|s\rangle $ acts on any
quantum state $|x\rangle |S\rangle $ a phase factor $\exp (-i\theta )$ is
generated only when the quantum state $|x\rangle =|s\rangle ,$ otherwise the
quantum state $|x\rangle |S\rangle $ keeps unchanged,

$\qquad \qquad \qquad C_{s}(\theta )|x\rangle |S\rangle =\exp (-i\theta
\delta _{sx})|x\rangle |S\rangle .\qquad \qquad \qquad \qquad \qquad (7)$ 
\newline
When the selective phase-shift operation $C_{s}(\theta )$ is applied to any
superposition $|\Psi \rangle =\stackrel{N-1}{\stackunder{x=0}{\sum }}%
a_{x}|x\rangle |S\rangle $ of the quantum system, the unitary evolution
process of the quantum system is just the same as Eq.(5) of the oracle
unitary operation $U_{o}(\theta )$,

$\qquad C_{s}(\theta )|\Psi \rangle =\exp (-i\theta )a_{s}|s\rangle
|S\rangle +\stackrel{N-1}{\stackunder{x=0,x\neq s}{\sum }}a_{x}|x\rangle
|S\rangle .\qquad \qquad \qquad \qquad (8)$ \newline
Therefore, the selective phase-shift operation $C_{s}(\theta )$ is really
equivalent to the oracle unitary operation $U_{o}(\theta )$ in the unitary
transformation and one can use exactly the selective phase-shift operation $%
C_{s}(\theta )$ to replace the oracle unitary operation $U_{o}$ to analyze
the unitary evolution process of the quantum system under the action of the
oracle unitary operation $U_{o}(\theta )$. However, it must be pointed out
that the selective phase-shift operation $C_{s}(\theta )$ is only applied to
the work qubits $I$ and independent of the auxiliary qubits $S$, while the
oracle unitary operation $U_{o}(\theta )$ is related to the whole quantum
system.

In order to describe conveniently the unitary evolution process of the
quantum system under the selective phase-shift operations one may use the
quantum state vector, i.e., the unity-number vector $\{a_{k}^{s}=\pm 1\}$
[17] to represent explicitly the selective phase-shift operation $%
C_{s}(\theta ).$ In the quantum state vector $\{a_{k}^{s}\}$ representation
the diagonal operator $D_{s}$ of the marked state can be expressed as

$D_{s}=(\frac{1}{2}E_{1}+a_{1}^{s}I_{1z})\bigotimes (\frac{1}{2}%
E_{2}+a_{2}^{s}I_{2z})\bigotimes ...\bigotimes (\frac{1}{2}%
E_{n}+a_{n}^{s}I_{nz}).\qquad \qquad \quad (9)$ \newline
It has been shown that the quantum state vector $\{a_{k}^{s}\}$ also
determines uniquely the quantum state $|s\rangle $ [17]$.$ Therefore, one
may design a quantum search algorithm or sequence to determine directly the
quantum state vector $\{a_{k}^{s}\}$ and then obtains further the marked
quantum state $|s\rangle $ instead of finding directly the marked quantum
state $|s\rangle $ by a quantum search algorithm$.$ This is different from
the conventional quantum search strategy [3, 11]. This is also the starting
point of the quantum search algorithms or sequences based on the quantum
mechanical unitary dynamics. Obviously, the oracle unitary operation $%
U_{o}(\theta )$ contains the desired information to determine certainly the
solution $|s\rangle $ to the quantum search problem. This also can be seen
from the fact that the oracle unitary operation $U_{o}(\theta )$ is
equivalent to the selective phase-shift operation $C_{s}(\theta )$, while
the latter is determined uniquely by the quantum state vector $\{a_{k}^{s}\}$%
. Now an NMR ensemble quantum computer operating on the highly mixed states
of a spin ensemble is used to solve the quantum search problem. As shown in
Eqs.(2), (3), (5), and (8) the oracle unitary operations $U_{f}$ and $%
U_{o}(\theta )$ are independent of any quantum state $|x\rangle $ of the
work qubits $I$ but are related to the quantum state $|S\rangle $ of the
auxiliary qubits $S$. When a quantum search algorithm based on the pure
states and built up with the oracle unitary operations and
oracle-independent unitary operations is used to determine the marked state $%
|s\rangle $ to the search problem on a spin ensemble, the quantum state $%
|S\rangle $ of the auxiliary qubits $S$ in the algorithm should be retained
or replaced by the effective pure state in the spin ensemble [12], but any
mixed states of the work qubits $I$ of the spin ensemble can be used as the
input states of the algorithm. For example, for the oracle unitary operation 
$U_{o}(\theta )$ the auxiliary quantum state $|S\rangle \langle S|$ $%
(|S\rangle =|0\rangle |1\rangle $ (see Eq.(3)) in the matrix representation
can be expressed as

$\qquad |S\rangle \langle S|=|0\rangle \langle 0|\bigotimes |1\rangle
\langle 1|=\frac{1}{4}E+\frac{1}{2}(S_{1z}-S_{2z})-S_{1z}S_{2z}.\qquad
\qquad \quad \ (10)$\newline
Then the auxiliary quantum state $|S\rangle \langle S|$ can be replaced by
the effective pure state taking the same form as Eq.(10) and the initial
quantum state $|x\rangle $ of the work qubits $I$ of the algorithm can be
replaced even by any highly mixed state of the work qubits $I$ in the spin
ensemble. However, the mixed state must be related to the marked state $%
|s\rangle $ in order that the quantum state vector $\{a_{k}^{s}\}$ of the
marked state $|s\rangle $ can be determined certainly by the algorithm. Note
that any magnetization operator $I_{k\mu }$ $(k=1,2,...,n;$ $\mu =x,y)$ and
the diagonal operator $D_{s}$ of the marked state do not commute$.$ It is
easy to prove that the following density operators $\rho (0)$ of the spin
ensemble always contains the marked state $|s\rangle $,

$\rho (0)=\rho _{I}(0)|S\rangle \langle S|=(\alpha E+\varepsilon _{k}I_{k\mu
})|S\rangle \langle S|$ $(k=1,2,...,n;$ $\mu =x,y),\quad (11a)$\newline
and

$\rho (0)=\rho _{I}(0)|S\rangle \langle S|=(\alpha E+\stackrel{n}{%
\stackunder{k=1}{\sum }}\varepsilon _{k}I_{k\mu })|S\rangle \langle S|$ $%
\qquad \qquad \qquad \qquad \ \ \quad \ \ (11b)$\newline
where the operator $E$ is unity operator, $\varepsilon _{k}$ is the spin
polarization parameter of the spin $k$ in the spin ensemble, and $\alpha $
is the normalized constant.

Now one can analyze conveniently the time evolution process of a quantum
spin system or its spin ensemble under the action of the oracle unitary
operation $U_{o}(\theta )$ by using the following general unitary
transformation that describes the action of the selective phase-shift
operation $C_{s}(\theta )$ on any density operator $\rho _{I}(0)$ of the
work qubits $I$ of the spin ensemble [28]:

$\qquad C_{s}(\theta )\rho _{I}(0)C_{s}(\theta )^{-1}=\rho _{I}(0)-(1-\cos
\theta )[\rho _{I}(0),D_{s}]_{+}$

$\qquad \qquad \qquad +i\sin \theta [\rho _{I}(0),D_{s}]+[(1-\cos \theta
)^{2}+\sin ^{2}\theta ]D_{s}\rho _{I}(0)D_{s}$ \qquad $\ (12)$ \newline
where the commutation $[\rho _{I}(0),D_{s}]_{+}=\rho _{I}(0)D_{s}+D_{s}\rho
_{I}(0)$ and $[\rho _{I}(0),D_{s}]=\rho _{I}(0)D_{s}-D_{s}\rho _{I}(0).$ A
general unitary transformation with a sequence of many selective phase-shift
operations $\{C_{k}(\theta _{k})\}$ can be derived directly from the basis
unitary transformation (12),

$U_{o}(\theta _{0},\theta _{1},...,\theta _{m-1})\rho _{I}(t)U_{o}(\theta
_{0},\theta _{1},...,\theta _{m-1})^{-1}$

$=\rho _{I}(t)-[\rho _{I}(t),$ $\stackrel{m-1}{\stackunder{k=0}{\sum }}%
(1-\cos \theta _{k})D_{k}]_{+}+i[\rho _{I}(t),$ $\stackrel{m-1}{\stackunder{%
k=0}{\sum }}D_{k}\sin \theta _{k}]$

$+$ $\stackrel{m-1}{\stackunder{k=0}{\sum }}\stackrel{m-1}{\stackunder{l=0}{%
\sum }}[(1-\cos \theta _{k})(1-\cos \theta _{l})+\sin \theta _{k}\sin \theta
_{l}]D_{k}\rho _{I}(t)D_{l}$

$+\stackrel{m-1}{\stackunder{l>k=0}{\sum }}i[\sin \theta _{k}(1-\cos \theta
_{l})-\sin \theta _{l}(1-\cos \theta _{k})](D_{k}\rho _{I}(t)D_{l}-D_{l}\rho
_{I}(t)D_{k})\ \quad (13)$ \newline
where the diagonal unitary operation $U_{o}(\theta _{0},\theta
_{1},...,\theta _{m-1})$ is a sequence of the selective phase-shift
operations:

$U_{o}(\theta _{0},\theta _{1},...,\theta _{m-1})=\stackrel{m-1}{\stackunder{%
k=0}{\prod }}C_{k}(\theta _{k}).$\newline
On the basis of the unitary transformation (12) one can design a simple
quantum search sequence to solve efficiently the quantum search problem on a
spin ensemble. First, the spin ensemble is prepared at the initial mixed
state of Eq.(11b). Then the oracle unitary operation $U_{o}(\theta )$ is
applied to the spin ensemble. The spin ensemble will evolve according to the
unitary transformation (12). This unitary transformation is given explicitly
below by inserting the diagonal operator $D_{s}$ of Eq.(9) in the quantum
state vector $\{a_{k}^{s}\}$ representation and the initial density operator
of Eq.(11b) with the index $\mu =y$ into Eq.(12),

$U_{o}(\theta )\stackrel{n}{\stackunder{k=1}{\sum }}\varepsilon
_{k}I_{ky}U_{o}(\theta )^{-1}=\stackrel{n}{\stackunder{k=1}{\sum }}%
\varepsilon _{k}I_{ky}-$ $(1-\cos \theta )\stackrel{n}{\stackunder{k=1}{\sum 
}}(\frac{1}{2}E_{1}+a_{1}^{s}I_{1z})\bigotimes ...$

$\bigotimes (\frac{1}{2}E_{k-1}+a_{k-1}^{s}I_{k-1z})\bigotimes (\varepsilon
_{k}I_{ky})\bigotimes (\frac{1}{2}E_{k+1}+a_{k+1}^{s}I_{k+1z})\bigotimes ...$

$\bigotimes (\frac{1}{2}E_{n}+a_{n}^{s}I_{nz})-\sin \theta \stackrel{n}{%
\stackunder{k=1}{\sum }}(\frac{1}{2}E_{1}+a_{1}^{s}I_{1z})\bigotimes
...\bigotimes (\frac{1}{2}E_{k-1}+a_{k-1}^{s}I_{k-1z})$

$\bigotimes (\varepsilon _{k}a_{k}^{s}I_{kx})\bigotimes (\frac{1}{2}%
E_{k+1}+a_{k+1}^{s}I_{k+1z})\bigotimes ...\bigotimes (\frac{1}{2}%
E_{n}+a_{n}^{s}I_{nz})\qquad \ \qquad \ \ \quad (14)$ \newline
where the term proportional to the unity operator $E$ and the auxiliary
quantum state $|S\rangle \langle S|$ of the initial density operator of
Eq.(11b) are omitted without losing generality$.$ Note that here the
auxiliary quantum state $|S\rangle \langle S|$ is taken as the effective
pure state and is of the $LOMSO$ operator subspace, and hence it will keep
unchanged when a $z-$direction gradient magnetic field is applied to the
spin ensemble [6, 7]. Obviously, the generated mixed state of Eq.(14) under
the action of the oracle unitary operation $U_{o}(\theta )$ on the spin
ensemble is quite complicated, but it can be simplified by using the
selective decoherence manipulation. First, a $90$ degree hard pulse along $%
y- $direction is applied to all the spins of the work qubits $I.$ Then a $z-$%
direction gradient magnetic field [6, 7] is applied to the spin ensemble to
cancel all the nonzero-order multiple-quantum coherence including the
single-quantum coherence, and then a zero-quantum dephasing pulse [29] is
applied to the spin ensemble to cancel all the zero-quantum coherence.. In
the final the density operator of Eq.(14) is reduced to the form

$U_{o}(\theta )\stackrel{n}{\stackunder{k=1}{\sum }}\varepsilon
_{k}I_{ky}U_{o}(\theta )^{-1}\Rightarrow \rho _{f}=(2/N)\stackrel{n}{%
\stackunder{k=1}{\sum }}\varepsilon _{k}a_{k}^{s}I_{kz}\qquad \qquad \qquad
\ \ \qquad (15)$ \newline
Obviously, the density operator $\rho _{f}$ of Eq.(15) carries the desired
information, i.e., the quantum state vector $\{a_{k}^{s}\}$, to solve the
quantum search problem$.$ Now, a read pulse $90_{-x}^{\circ }$ is applied to
all the spins of the work qubits $I$ and then the NMR receiver is turned on
to record in phase-sensitive mode the NMR signal of the density operator $%
\rho _{f}$ of Eq.(15). In order to cancel the effect of the auxiliary qubits 
$S$ of Eq.(10) on the recorded NMR signal the spins of the auxiliary qubits $%
S$ are decoupled during recording the NMR signal. From the fast Fourier
transform spectrum of the recorded NMR signal one can determine certainly
the quantum state vector $\{a_{k}^{s}\}$ when the signal-to-noise ratio of
the NMR signal of Eq.(15) is high enough. The density operator of Eq.(15)
shows that the NMR signal intensity is proportional to the factor $(2/N)$,
indicating that the NMR signal intensity reduces exponentially as the qubit
number of the work qubits $I$. This characteristic feature is similar to the
classical NMR quantum computation based on the effective pure states [15].
Therefore, the above quantum search algorithm can only be used to solve
efficiently the quantum search problem with a few qubits.

However, the significant difference between the quantum search algorithms
based on the spin ensemble and the pure quantum state [3] or the effective
pure state [11] is that the Grover quantum search algorithm based on the
pure quantum states [3] or the effective pure state [11] can achieve only
quadratic speedup over the classical counterparts due to the limit of
quantum mechanical measurement [6], which is a limit in quantum mechanical
principle, that is, the conventional Grover algorithm is a quadratic but not
polynomial-time algorithm, but the ensemble quantum search sequence can
achieve an exponential-time speedup over the classical counterparts, that
is, this sequence is a polynomial-time one (one performs only twice times
the oracle unitary operation $U_{f}$ in the algorithm) and really moves the
frontier between the solvability and intractability of complexity problems.
Though the final output NMR signal of the ensemble quantum search sequence
reduces exponentially as the size of the solved search problem, this limit
is in techniques but not in quantum mechanical principles and therefore,
this limit could be overcome as the relevant techniques are improved [7, 8,
22, 23, 24]. There is also another advantage for the ensemble quantum search
sequence that the sequence can be run even on a spin system with a short
decoherence time since in the whole sequence the oracle unitary operation $%
U_{f}$ is performed only twice times.

There may be two possible schemes [18] to improve the above ensemble quantum
search sequence so that it could be used to solve the search problem with a
larger number of qubits. One possible scheme is that the initial spin
polarization of a spin ensemble is enhanced sufficiently with various
possible NMR signal enhancement techniques including spin polarization
transfer techniques [6, 7, 22] and other techniques [23, 24] so that the NMR
signal of the density operator of Eq.(15) still can be observable even in
the spin system with a larger number of qubits, although the final NMR
signal of the sequence is reduced exponentially as the qubit number of the
spin system. This scheme retains the original advantage of the sequence that
it still can be run on the spin systems with a short decoherence time.
However, this scheme is rigorous in technique. Another scheme is that one
could enhance the final detectable NMR signal by running a quantum search
sequence that consists of a number of the oracle unitary operations and
oracle-independent unitary operations on the spin ensemble. But this scheme
requires that the spin ensemble must have a long decoherence time.
Obviously, there is a maximum call number $M_{s}$ of the oracle unitary
operation $U_{o}(\theta )$ or a maximum running time $T_{2\max }$ of the
quantum search sequence in a spin ensemble due to the decoherence effect.
The final output NMR signal of the sequence will decrease simply but not be
enhanced again as increasing the call number of the oracle unitary operation
in the sequence if the call number is larger than the maximum one $M_{s}$.
This is because in that case running the sequence will take more time than $%
T_{2\max }$, so that the decoherence effect becomes the dominant effect
negatively upon the final NMR signal.

A simple method to enhance the output NMR signal by many calls of the oracle
unitary operation is the conventional spin echo experiments [5, 7]. The
qubit interaction in the diagonal operator $D_{s}$ of the marked state $%
|s\rangle $ can be decoupled one by one by the spin-echo experiments. First,
any Hamiltonian $D_{s}(n-k)$ $(k=1,2,...,n-1)$ that is independent of $k$
qubits of the spin system is prepared with the spin echo sequences by
starting the diagonal operator $D_{s}$,

$D_{s}(n-1)=D_{s}+\exp (-i\pi I_{nx})D_{s}\exp (i\pi I_{nx}),$

$D_{s}(n-2)=D_{s}(n-1)+\exp (-i\pi I_{n-1x})D_{s}(n-1)\exp (i\pi
I_{n-1x}),...,$

$D_{s}(n-k)=D_{s}(n-k+1)+\exp (-i\pi I_{n-k+1x})D_{s}(n-k+1)\exp (i\pi
I_{n-k+1x}),$\newline
where $D_{s}(n-k)\equiv \stackrel{n-k}{\stackunder{l=1}{\bigotimes }}(\frac{1%
}{2}E_{l}+a_{l}^{s}I_{lz}).$ Then one builds up a new auxiliary oracle
unitary operation\ $V_{os}(k,\theta )=\exp [-i\theta D_{s}(n-k)]$ with the
Hamiltonian $D_{s}(n-k).$ There are $2^{k}$ oracle unitary operations $%
U_{o}(\theta )$ in the unitary transformation $V_{os}(k,\theta )$. Now using
the oracle unitary operation $V_{os}(k,\theta )$ to replace the oracle
unitary operation $U_{o}(\theta )$ in Eq.(14) to perform quantum search one
can find that the final NMR signal intensity of Eq.(15) is proportional to
the factor $2^{k},$ indicating that the output NMR signal intensity of the
new quantum search sequence increases linearly as the call number of the
oracle unitary operation. Obviously, this is an ineffective sequence to
extend the realizable size of the NMR\ ensemble quantum computation. An
improved approach to the simple spin-echo experiments is that following the
action of the unitary operation $V_{os}(k,\theta )$ on the initial density
operator one applies a sequence of non-oracle selective unitary operations $%
\stackunder{r}{\prod }C_{r}(\theta )$ to the spin ensemble. Then the NMR
signal intensity of Eq.(15) is proportional to the sum $\stackunder{k}{\sum }%
\varepsilon _{k}(\stackunder{s}{\sum }a_{k}^{s}+\stackunder{r}{\sum }%
a_{k}^{r})I_{kx}$, where the quantum state vectors $\{a_{k}^{s}\}$ and $%
\{a_{k}^{r}\}$ are related to the oracle and the non-oracle selective
unitary operations, respectively. Now, if the factors $\stackunder{s}{\sum }%
a_{k}^{s}$ and $\stackunder{r}{\sum }a_{k}^{r}$ have the same sign then the
final NMR\ signal is enhanced, otherwise it is reduced. Therefore, the NMR
signal intensity may be enhanced when the non-oracle selective unitary
operations are chosen properly. The more the number of the non-oracle
selective unitary operations is in the sequence, the stronger the NMR signal
is. However, there is a question that how many non-oracle selective unitary
operations can be used to enhance effectively the NMR signal under the
condition that the quantum state vector $\{a_{k}^{s}\}$ can be extracted
certainly from the final NMR signal intensity. This question limits really
the maximum enhancement of the NMR signal by the sequence since the more the
non-oracle selective unitary operations are used in the sequence it becomes
more difficult for the solution information of the search problem to be
extracted from the NMR signal. In order to improve further the method the
NMR multiple-quantum spectroscopy is proposed to solve the search problem in
the following. I will discuss the quantum search NMR multiple-quantum
spectroscopy from the view point of the NMR experiments and leave its
complexity problem in the future. \newline
\[
\]
\newline
{\Large 3. The NMR multiple-quantum spectroscopy}

The basic idea for the multiple-quantum NMR spectroscopic method used to
perform the quantum search is that information of the unknown marked state $%
|s\rangle $ is first loaded on the multiple-quantum coherence of a spin
ensemble by using iterative sequences of the oracle unitary operations $%
U_{o}(\theta )$ and oracle-independent unitary operations, the search
problem then could be solved by extracting the desired information of the
marked state from these NMR multiple-quantum transition spectra obtained by
NMR measurement techniques. The important point for the method is that the
multiple-quantum coherence of the spin ensemble need first to be excited
effectively by a sequence of the oracle unitary operations and
oracle-independent unitary operations. Here a general (formal) quantum
search sequence based on the multiple-quantum NMR spectroscopy [7, 8] is
proposed. The sequence is constructed with the oracle unitary operations and
oracle-independent unitary operations and may be divided into the following
several steps:\newline
(a) The suitable initial input state for the quantum search sequence is
prepared at the mixed state of a spin ensemble such as Eq.(11a) or (11b),

$\rho (0)=\rho _{I}(0)|S\rangle \langle S|=(\alpha _{0}E+\stackrel{n}{%
\stackunder{k=1}{\sum }}\varepsilon _{k}I_{kp})|S\rangle \langle S|$ $%
(p=x,y,z).$\newline
(b) Apply the iterative sequence $U(D_{s})$ which is built up with the
oracle unitary operations $U_{o}(\theta )$ and oracle-independent unitary
operations to the initial mixed state $\rho (0)$ to excite the
multiple-quantum coherence,

$\rho (\tau _{1})=U(D_{s})\rho (0)U(D_{s})^{+}$\newline
(c) During the $t_{1}$ evolution period under the effective spin Hamiltonian 
$H$ the multiple-quantum coherence are labelled by their own frequencies so
that various order multiple-quantum coherence can be distinguished clearly
in NMR multiple-quantum spectra,

$\rho (t_{1},\tau _{1})=\exp (-iHt_{1})\rho (\tau _{1})\exp (iHt_{1})$ 
\newline
(d) Apply the second iterative sequence $V(D_{s})$ of the oracle unitary
operations and oracle-independent unitary operations to convert the
multiple-quantum coherence into the single-quantum coherence so that they
can be detected by NMR\ probes.

$\rho (\tau _{2},t_{1,}\tau _{1})=V(D_{s})\rho (t_{1,}\tau _{1})V(D_{s})^{+}$%
\newline
(e) The NMR\ signal detection along with $q-$direction $(q=x,y,z)$

$\rho _{f}=\rho _{f}(t_{1})=Tr\{\stackrel{n}{\stackunder{k=1}{\sum }}%
I_{kq}\rho (\tau _{2},t_{1,}\tau _{1})\}.$\newline
The longitudinal component $I_{kz}$ in the final NMR signal $\rho _{f}$ is
not observable in NMR experiments, but it becomes observable if an
additional $90$ degree pulse converts it into the single-quantum coherence.
The multiple-quantum coherence are usually detected indirectly by NMR
measurement techniques such as the two-dimensional NMR method [7, 8]. The
usual two-dimensional NMR experiment methods have been developed to do
quantum computation on the basis of the concept of the effective pure states
[12], but here NMR multiple-quantum spectroscopy is emphasized. The oracle
iterative sequence $U(D_{s})$ carries the information of the marked state $%
|s\rangle $ and in the step (b) the information is really transferred into
the multiple-quantum coherence of the spin ensemble through applying the
iterative sequence $U(D_{s})$ to the initial density operator of the spin
ensemble. The multiple-quantum coherence (or transitions) are detected
indirectly by the two-dimensional NMR measurement techniques [7, 8], which
is achieved by the steps (c), (d) and (e) in the sequence. Obviously, the
output NMR signal $\rho _{f}(t_{1})$ carries the information of the marked
state $|s\rangle $. Then by fast Fourier transforming the output NMR signal $%
\rho _{f}(t_{1})$ on the time variable $t_{1}$ one obtains the
multiple-quantum spectrum. Now the spectral parameters including the
multiple-quantum resonance frequencies, phase, and intensities of the
multiple-quantum spectrum contain the information of the marked state $%
|s\rangle $. Then one could extract the quantum state vector $\{a_{k}^{s}\}$
and hence find the marked state $|s\rangle $ from these spectral parameters
by NMR data processing$.$ \newline
\newline
\textbf{3.1 The expression for the output NMR signal }

For convenient discussion, here consider a homonuclear spin ensemble where
the spin polarization parameter $\varepsilon _{k}$ for all the spins in the
ensemble are the same $(\varepsilon =\varepsilon _{k})$. Then the initial
density operator of work qubits $I$ in the step (a) is written briefly as $%
\rho _{I}(0)=\varepsilon F_{p}=\varepsilon \stackrel{n}{\stackunder{k=1}{%
\sum }}I_{kp}$ without losing generality since the auxiliary quantum state $%
|S\rangle \langle S|,$ the term proportional to the unity operator $E,$ and
the spin polarization factor $\varepsilon _{k}$ keep unchanged during the
evolution process of the quantum search sequence. From the steps (a)-(e) of
the sequence the final NMR signal intensity of the sequence can be generally
expressed as

$\rho _{f}(t_{1})=Tr\{F_{q}V(D_{s})\exp (-iHt_{1})U(D_{s})\varepsilon
F_{p}U(D_{s})^{+}\exp (iHt_{1})V(D_{s})^{+}\}$

$=\varepsilon Tr\{V(D_{s})^{+}F_{q}V(D_{s})\exp
(-iHt_{1})U(D_{s})F_{p}U(D_{s})^{+}\exp (iHt_{1})\}\qquad (16)$\newline
where the invariance of the trace operation $Tr$ to cyclic permutation has
been introduced. In high-resolution NMR spectroscopy the spin Hamiltonian $H$
of the spin system in the spin ensemble in a high magnetic field is a
longitudinal magnetization and spin order $(LOMSO)$ operator and usually can
be written as [7, 8]

$\quad \qquad \qquad \qquad H=\stackrel{n}{\stackunder{k=1}{\sum }}\Omega
_{k}I_{kz}+\stackrel{n}{\stackunder{k>l}{\sum }}2\pi
J_{kl}I_{kz}I_{lz}\qquad \qquad \qquad \qquad \qquad (17)$ \newline
With the help of the eigenbase $|k\rangle $ of the Hamiltonian $H$, $%
H|k\rangle =\omega _{k}|k\rangle \ (\hslash =1),$ which are also the usual
quantum computational bases, the NMR signal of Eq.(16) can be expanded as

$\qquad \qquad \qquad \rho _{f}(t_{1})=\varepsilon \stackunder{j,k}{%
\stackrel{}{\sum }}Q_{jk}^{*}P_{jk}\exp (-i\omega _{jk}t_{1})\qquad \qquad
\qquad \quad \qquad \ (18)$ \newline
where the multiple-quantum transition frequency $\omega _{jk}=\omega
_{j}-\omega _{k}$ and the Hermitian operators $P(D_{s})$ and $Q(D_{s})$ are
defined as

$\qquad \qquad P(D_{s})=U(D_{s})F_{p}U(D_{s})^{+},\qquad \qquad \qquad \
\qquad \qquad \qquad \quad \ \ (19a)$

$\qquad \qquad Q(D_{s})^{+}=Q(D_{s})=V(D_{s})^{+}F_{q}V(D_{s}),\qquad \qquad
\qquad \qquad \quad (19b)$ \newline
and the matrix elements $P_{jk}$ and $Q_{jk}$ are given by

$\qquad \qquad P_{jk}=\langle j|P(D_{s})|k\rangle ,$ $Q_{jk}=\langle
j|Q(D_{s})|k\rangle .$ \newline
Although the eigenvalue $\omega _{k}$ of the Hamiltonian $H$ is independent
of the quantum state vector $\{a_{k}^{s}\}$ of the marked state $|s\rangle $%
, the excited multiple-quantum transition frequencies $\{\omega _{jk}\}$ may
contain the information of the marked state $|s\rangle $ besides the
amplitudes and phases of the multiple-quantum transitions since the
excitation of the multiple-quantum coherence by the oracle unitary sequence $%
U(D_{s})$ is dependent on the marked state. Actually, the oracle unitary
sequence $U(D_{s})$ decides which orders multiple-quantum coherence are
excited effectively and finally detected in the quantum search sequence. If
the inphase multiple-quantum spectrum is required to be observed for a given
quantum order one had better require that the oracle unitary sequence $%
V(D_{s})$ satisfies the relation [30]:

$\qquad \qquad \qquad Q(D_{s})^{+}=\exp (-i\varphi F_{z})P(D_{s})\exp
(i\varphi F_{z}).\qquad \qquad \qquad \ \ (20)$ \newline
Then the NMR signal intensity of Eq.(18) can be further expressed as

$\qquad \qquad \rho _{f}(t_{1})=\varepsilon \stackunder{m}{\stackrel{}{\sum }%
}\stackunder{j,k}{\stackrel{}{\sum }}|P_{jk}|^{2}\exp (im\varphi )\exp
(-i\omega _{jk}t_{1})\qquad \qquad \qquad \ (21)$ \newline
where the quantum order $m=M_{j}-M_{k}$ and $M_{k}$ is the magnetic quantum
number: $F_{z}|k\rangle =M_{k}|k\rangle $ $(\hslash =1).$ The NMR signal of
Eq.(21) indicates that each $m-$order multiple-quantum transition has the
same phase, although the multiple-quantum transitions with different quantum
order $m$ may have a phase difference. Therefore, by using separation
techniques for different order multiple-quantum transitions such as TPPI
technique [7, 8] one can record the inphase multiple-quantum peaks with
their own resonance frequencies or quantum order in the multiple-quantum
spectrum. The intensities, frequencies, and phases of the multiple-quantum
peaks carry the information of the marked state. Then from these spectral
parameters one could extract the desired information of the marked state.
Although the inphase multiple-quantum spectrum of Eq.(21) may be more
convenient and useful in the measurement of structural properties of
molecules in NMR multiple-quantum spectroscopy [30], the multiple-quantum
spectrum of Eq.(18) may contain richer information of the marked state and
may be more useful to solve the quantum search problem.

If the effective spin Hamiltonian is prepared as $H=\omega F_{z}$ instead of
that one of Eq.(17) in the $t_{1}$ evolution period, all the
multiple-quantum coherence with the same quantum order have the same
precession frequency. That is, for all the $k-$order multiple-quantum
coherence their precession frequencies are the same and equal $f_{k}=k\omega 
$ $(k=0,\pm 1,\pm 2,...,\pm n)$. In this case there are $(2n+1)$ different
order multiple-quantum resonance peaks with their own resonance frequencies $%
\{f_{k}\}$ in the multiple-quantum spectrum for an $n-$qubit spin system.
Since different order multiple-quantum coherence in the density operator $%
\rho _{f}(t_{1})$ of Eq.(18) are labelled with their own transition
frequencies (or precession frequencies), they can be distinguished in the
NMR multiple-quantum frequency spectrum. It is known that number of
independent multiple-quantum coherence increases exponentially as the qubit
number in a spin system [7]. If each multiple-quantum coherence has its own
precession frequency in a spin system, then amplitude of each
multiple-quantum coherence will decrease exponentially as the qubit number
and this will make each multiple-quantum spectral peak unobservable in the
multiple-quantum spectrum of a spin system with many qubits since the total
amplitude of the multiple-quantum coherence of a spin system is fixed and is
determined by the initial density operator. The importance to choose the
spin Hamiltonian $H=\omega F_{z}$ to label the multiple-quantum coherence in
the $t_{1}$ evolution period instead of the spin Hamiltonian of Eq.(17)
therefore becomes very clear: all the same order multiple-quantum coherence
have the same precession frequency and their resonance frequency peaks in
the multiple-quantum spectrum can be coherently co-added in amplitude and
become a single spectral peak, as can be seen in the NMR signal of Eq.(18),
so that the multiple-quantum spectral peak may increase greatly in a
favorable case. This shows that at least some lower order multiple-quantum
spectral peaks in the spin ensemble do not reduce exponentially as the qubit
number if the initial density operator is excited effectively to the
multiple-quantum coherence by the oracle iterative sequence $U(D_{s})$.
Therefore, complexity of the quantum search sequence now is only dependent
on the excitation of the multiple-quantum coherence by the oracle iterative
sequence $U(D_{s})$ and the extraction of the solution information of the
search problem by the NMR data processing. Of course, it could become more
difficult to extract the information of solution of the search problem using
this method labelling multiple-quantum coherence with respect to that one
using the Hamiltonian of Eq.(17). By the NMR signal expression (18) one can
explicitly calculate the output NMR signal intensities of various order
multiple-quantum transitions,

$\qquad \qquad \rho _{f}(t_{1})=\stackunder{m=-n}{\stackrel{n}{\sum }}%
[\varepsilon \stackunder{j,k}{\stackrel{}{\sum^{\prime }}}%
Q_{jk}^{*}P_{jk}]\exp (-im\omega t_{1})\qquad \qquad \qquad \qquad \ (22)$ 
\newline
where the sum $\sum^{\prime }$ runs only those indexes $(j,k)$ with the
quantum order $m=M_{j}-M_{k}.$ The $m-$order multiple-quantum peak intensity
and phase therefore are determined from

$\qquad \qquad \qquad I(m)=[\varepsilon \stackunder{j,k}{\stackrel{}{%
\sum^{\prime }}}Q_{jk}^{*}P_{jk}].\qquad \qquad \qquad \qquad \qquad \qquad
\qquad \ (23)$ \newline
Now the information of solution of the search problem could be extracted
from the spectral parameters $I(m)$ by NMR data processing. \newline
\newline
\textbf{3.2 The multiple-quantum operator algebra spaces and the
construction of the oracle unitary sequences}

The multiple-quantum operator algebra space [24, 25] states that the whole
Liouville operator space of a spin ensemble contains the even-order
multiple-quantum operator subspace, the zero-quantum operator subspace, and
the longitudinal magnetization and spin order ($LOMSO$) operator subspace
and each former subspace contains the followed subspaces. For example, the
zero-quantum operator subspace contains the whole $LOMSO$ operator subspace.
There are important properties for the zero-quantum operators and the
even-order multiple-quantum operators. Any operator of the Liouville
operator space will keep its quantum order unchanged when it is acted on by
a zero-quantum operator and the parity of quantum order of an operator of
the Liouville operator space can not be changed when the operator is acted
on by an even-order multiple-quantum operator [24, 31]. With the help of the
closed properties of these operator subspaces one may simplify the
construction of the oracle unitary sequences $U(D_{s})$ and $V(D_{s})$. In
general, the oracle unitary sequences $U(D_{s})$ and $V(D_{s})$ are composed
of the oracle unitary operation $U_{o}(\theta )$ and oracle-independent
unitary operations. Suppose that the oracle unitary sequence $U(D_{s})$ is
constructed in the form:

$\qquad \qquad \qquad U(D_{s})=R_{k}O_{k-1}R_{k-1}...O_{1}R_{1}\qquad \qquad
\qquad \quad \qquad \qquad \ (24)$ \newline
where each unitary operation $R_{k}$ is an oracle-independent unitary
operation and $O_{k}=O_{k}(D_{s})$ an oracle unitary operation sequence
which contains the oracle unitary operation $U_{o}(\theta ).$ Now assume
that the initial mixed state $\rho _{I}(0)$ in the quantum search sequence
is taken as the longitudinal magnetization operator, $\rho _{I}(0)=\stackrel{%
n}{\stackunder{k=1}{\sum }}\varepsilon _{k}I_{kz},$ which is also a
zero-quantum operator. There may be two simpler schemes to load the
information of the marked state $|s\rangle $ on the multiple-quantum
spectrum. One scheme is that only the zero-quantum transitions are used to
carry the information of the marked state. The zero-quantum operator
subspace is the smaller operator subspace in the Liouville operator space.
Then it is a better scheme to load the information of the marked state on
the zero-quantum operator subspace since this could simplify the loading and
extraction of the desired information of the marked state. Obviously, this
requires that the oracle unitary sequence $U(D_{s})$ should be prepared as a
zero-quantum unitary operator. When the oracle unitary sequence $U(D_{s})$
is applied to the initial mixed state $\rho _{I}(0)$ only the zero-quantum
coherence of the spin system can be excited according to closed property of
the multiple-quantum operator algebra space since the initial mixed state $%
\rho _{I}(0)$ is also a zero-quantum operator. In order to distinguish the
effects between the oracle unitary operations $\{O_{k}\}$ and the
oracle-independent unitary operations $\{R_{k}\}$ on the excitation of the
zero-quantum coherence it could be better to choose the unitary operations $%
\{R_{k}\}$ as the $LOMSO$ unitary operators and the oracle unitary
operations $\{O_{k}\}$ as the zero-quantum unitary operators. It is known
from multiple-quantum operator algebra space formalism that any pair of the $%
LOMSO$ operators commute with each other, and the closed property of the
zero-quantum operator subspace shows that any zero quantum operator is
transformed unitarily into a zero-quantum operator when it is acted on by
any zero-quantum unitary operator. Then the zero-quantum coherence will be
excited only by the oracle unitary operations $\{O_{k}\}$, while the $LOMSO$
unitary operations $\{R_{k}\}$ should have not net contribution to the
excitation of the zero-quantum coherence.. But the unitary operators $%
\{R_{k}\}$ could be chosen properly so that the excitation efficiency of the
zero-quantum coherence can be enhanced. If the zero-quantum coherence
spectra are observed in the output NMR signal of the quantum search sequence
it can be sure that the zero-quantum spectra contain the information of the
marked state. Another scheme is that the nonzero order multiple-quantum
coherence and especially the even-order multiple-quantum coherence are used
to carry the information of the marked state. According to the
multiple-quantum operator algebra space formalism all the even-order
multiple-quantum operators form a closed operator subspace of the Liouville
operator space. The closed property of the even order multiple-quantum
operator subspace shows that the initial density operator (a $LOMSO$
operator) will be transferred into even order multiple-quantum coherence if
it is acted on by an even-order multiple-quantum unitary operator.
Therefore, if all the oracle unitary operations $\{O_{k}\}$ are taken as the
even-order multiple-quantum unitary operators and the oracle-independent
unitary operations $\{R_{k}\}$ as the proper $LOMSO$ or zero-quantum unitary
operators, then even order multiple-quantum coherence will be excited by the
oracle unitary sequence $U(D_{s})$ in the spin ensemble and moreover, the
nonzero even-order multiple-quantum coherence can only be excited by the
oracle unitary operations $\{O_{k}\}$ and the oracle-independent unitary
operations $\{R_{k}\}$ will have not net contribution to the excitation of
the nonzero-order multiple-quantum coherence when the oracle unitary
sequence $U(D_{s})$ is applied to the initial density operator $\rho _{I}(0)$%
. Therefore, the information of the marked state now is contained mainly in
the nonzero even order multiple-quantum spectra of the output NMR signal of
the sequence. How easy to extract the information $\{a_{k}^{s}\}$ of the
marked state from these multiple-quantum spectra? This depends on what the
explicit forms of the oracle-independent operators $\{R_{k}\}$ and the
oracle unitary operations $\{O_{k}\}$ are taken. The importance is that the
oracle unitary sequence $U(D_{s})$ needs to be constructed properly so that
the multiple-quantum coherence can be excited effectively.\newline
\newline
\textbf{(a)\ the preparation for the zero-quantum unitary operators}

There are a number of methods to construct the zero-quantum unitary
operations with a sequence of the oracle unitary operations $U_{o}(\theta )$
and oracle-independent unitary operations. One simple and systematic method
is that the preparation can be performed by the discrete Fourier analysis
and the phase cycling technique [17] by starting the diagonal operator $%
D_{s} $ of the marked state $|s\rangle $. According to the multiple-quantum
operator algebra space any diagonal operator $D_{l},$ including the diagonal
operator $D_{s}$ of the marked state $|s\rangle ,$ can be expressed as a
linear combination of the $LOMSO$ basis product operators $\{Z_{k}\}=\{E,$ $%
I_{kz},$ $2I_{kz}I_{lz},$ $4I_{kz}I_{lz}I_{mz},...,$ $%
2^{n-1}I_{1z}I_{2z}...I_{nz}\}$ including the unity operator $E$ $(E=Z_{0})$
[24]$,$

$\qquad \qquad \qquad \qquad D_{k}=\stackrel{N-1}{\stackunder{l=0}{\sum }}%
(A)_{kl}Z_{l}.\qquad \qquad \qquad \qquad \qquad \quad \qquad \quad \ \ (25)$
\newline
Then the Hermitian product operators $X_{l}=\exp (-i\frac{\pi }{2}%
F_{y})Z_{l}\exp (i\frac{\pi }{2}F_{y})$ can be constructed by

$X_{l}=2^{l-1}I_{k_{1}x}I_{k_{2}x}...I_{k_{l}x}=\exp (-i\frac{\pi }{2}F_{y})[%
\stackrel{N-1}{\stackunder{j=0}{\sum }}(A^{-1})_{lj}D_{j}]\exp (i\frac{\pi }{%
2}F_{y})\qquad \ (26)$ \newline
and in a more general case any linear operator function $f(D_{s})$ related
to the diagonal operator $D_{s}$ of the marked state $|s\rangle $ and any
other diagonal operators $D_{r}$ by

$f(D_{s})=\stackrel{}{\stackunder{l=1}{\sum }}b_{l}X_{l}$

$=\exp (-i\frac{\pi }{2}F_{y})[\stackrel{}{\stackunder{l=1}{\sum }}b_{l}%
\stackrel{N-1}{\stackunder{j=0}{\sum }}(A^{-1})_{lj}D_{j}]\exp (i\frac{\pi }{%
2}F_{y})\qquad \qquad \qquad \qquad \qquad (27)$ \newline
where $b_{l}$ is a coefficient. Now by starting the operator function of
Eq.(27) one can prepare the Hermitian zero-quantum operators with aid of the
discrete Fourier analysis and the phase cycling technique [17],

$H_{zq}(D_{s})=\frac{1}{N_{1}}\stackrel{N_{1}-1}{\stackunder{k=0}{\sum }}%
\exp (-i\varphi _{k}F_{z})f(D_{s})\exp (-i\varphi _{k}F_{z})$

$=\frac{1}{N}\stackrel{n}{\stackunder{l>k=1}{\sum }}J_{kl}(\{a_{k}^{s}%
\})(I_{k}^{+}I_{l}^{-}+I_{k}^{-}I_{l}^{+})$

$+\frac{1}{N}\stackrel{n}{\stackunder{q>p>l>k=1}{\sum }}\{J_{klpq}(%
\{a_{k}^{s}%
\})(I_{k}^{+}I_{l}^{+}I_{p}^{-}I_{q}^{-}+I_{k}^{-}I_{l}^{-}I_{p}^{+}I_{q}^{+}) 
$

$+J_{klpq}^{\prime
}(\{a_{k}^{s}%
\})(I_{k}^{+}I_{l}^{-}I_{p}^{+}I_{q}^{-}+I_{k}^{-}I_{l}^{+}I_{p}^{-}I_{q}^{+}) 
$

$+J_{klpq}^{\prime \prime
}(\{a_{k}^{s}%
\})(I_{k}^{+}I_{l}^{-}I_{p}^{-}I_{q}^{+}+I_{k}^{-}I_{l}^{+}I_{p}^{+}I_{q}^{-})\}+...,\qquad \qquad \qquad \qquad \ \qquad \ \ \ (28) 
$ \newline
where the phase $\varphi _{k}=k2\pi /N_{1},$ $k=0,1,2,...,N_{1}-1$ and the
integer $N_{1}\geq n.$ By the $N_{1}-$step phase cycling systematically only
the zero-quantum operator $H_{zq}(D_{s})$ of the operator function $f(D_{s})$
is retained but all nonzero order multiple-quantum operators are cancelled.
The zero-quantum unitary operator then can be constructed generally with the
Hermitian zero-quantum operator of Eq.(28),

$\qquad \qquad \qquad U_{zq}(D_{s},t)=\exp [-iH_{zq}(D_{s})t].\qquad \qquad
\qquad \quad \qquad \qquad (29)$ \newline
Since the operator function $f(D_{s})$ contains the diagonal operator $D_{s}$
of the marked state $|s\rangle $ the zero-quantum operator $H_{zq}(D_{s})$
of Eq.(28) carries the information of the marked state $|s\rangle $. The
zero-quantum unitary operator of Eq.(29) may be further decomposed into a
sequence of the oracle unitary operations and oracle-independent unitary
operations by using the Trotter-Suzuki formula [32, 33]:

$\qquad \stackunder{m\rightarrow \infty }{\stackrel{}{\lim }}%
(e^{-iA_{1}/m}e^{-iA_{2}/m}...e^{-iA_{n}/m})^{m}=e^{-i(A_{1}+A_{2}+...+A_{n})}.\qquad \qquad \quad (30) 
$ \newline
Now the decomposition (30) of the zero-quantum unitary operator of Eq.(29)
may be used as the oracle unitary sequence $U(D_{s})$ to excite the
zero-quantum coherence in the quantum search sequence. \newline
\newline
\textbf{(b)\ the preparation for the nonzero-order multiple-quantum unitary
operators}

Using the discrete Fourier analysis and the phase cycling technique [17] by
starting the diagonal operator $D_{s}$ of the marked state $|s\rangle $ one
still can construct generally any nonzero order multiple-quantum coherence
operators and further uses the operators to construct the multiple-quantum
unitary operators. But a simple scheme to prepare a nonzero-order
multiple-quantum unitary operator as a sequence of the oracle unitary
operations and oracle-independent unitary operations is given below. For any
Hermitian product operator $X_{l}$ $=$ $%
2^{l-1}I_{k_{1}x}I_{k_{2}x}...I_{k_{l}x}$ of the $x-$direction product
operator set $\{E,$ $I_{kx},$ $2I_{kx}I_{lx},$ $4I_{kx}I_{lx}I_{mx},...\}$
one may construct a number of the nonzero order multiple-quantum operators
such as $i[X_{l},(D_{0}\pm D_{N-1})]_{\pm }$. The typical two of which are
given explicitly by

$i[X_{l},(D_{0}-D_{N-1})]$

$=+\frac{1}{2}i\stackrel{l}{\stackunder{j=1}{\bigotimes }}%
(I_{k_{j}x}-iI_{k_{j}y})\bigotimes (\frac{1}{2}E_{k_{l+1}}+I_{k_{l+1}z})%
\bigotimes ...\bigotimes (\frac{1}{2}E_{k_{n}}+I_{k_{n}z})$

$-\frac{1}{2}i\stackrel{l}{\stackunder{j=1}{\bigotimes }}%
(I_{k_{j}x}+iI_{k_{j}y})\bigotimes (\frac{1}{2}E_{k_{l+1}}-I_{k_{l+1}z})%
\bigotimes ...\bigotimes (\frac{1}{2}E_{k_{n}}-I_{k_{n}z})\qquad \qquad $

$-\frac{1}{2}i\stackrel{l}{\stackunder{j=1}{\bigotimes }}%
(I_{k_{j}x}+iI_{k_{j}y})\bigotimes (\frac{1}{2}E_{k_{l+1}}+I_{k_{l+1}z})%
\bigotimes ...\bigotimes (\frac{1}{2}E_{k_{n}}+I_{k_{n}z})$

$+\frac{1}{2}i\stackrel{l}{\stackunder{j=1}{\bigotimes }}%
(I_{k_{j}x}-iI_{k_{j}y})\bigotimes (\frac{1}{2}E_{k_{l+1}}-I_{k_{l+1}z})%
\bigotimes ...\bigotimes (\frac{1}{2}E_{k_{n}}-I_{k_{n}z}),\qquad \ \ \ \
(31)$

$[X_{l},(D_{0}+D_{N-1})]_{+}$

$=+\frac{1}{2}\stackrel{l}{\stackunder{j=1}{\bigotimes }}%
(I_{k_{j}x}-iI_{k_{j}y})\bigotimes (\frac{1}{2}E_{k_{l+1}}+I_{k_{l+1}z})%
\bigotimes ...\bigotimes (\frac{1}{2}E_{k_{n}}+I_{k_{n}z})$

$+\frac{1}{2}\stackrel{l}{\stackunder{j=1}{\bigotimes }}%
(I_{k_{j}x}+iI_{k_{j}y})\bigotimes (\frac{1}{2}E_{k_{l+1}}-I_{k_{l+1}z})%
\bigotimes ...\bigotimes (\frac{1}{2}E_{k_{n}}-I_{k_{n}z})$

$+\frac{1}{2}\stackrel{l}{\stackunder{j=1}{\bigotimes }}%
(I_{k_{j}x}+iI_{k_{j}y})\bigotimes (\frac{1}{2}E_{k_{l+1}}+I_{k_{l+1}z})%
\bigotimes ...\bigotimes (\frac{1}{2}E_{k_{n}}+I_{k_{n}z})$

$+\frac{1}{2}\stackrel{l}{\stackunder{j=1}{\bigotimes }}%
(I_{k_{j}x}-iI_{k_{j}y})\bigotimes (\frac{1}{2}E_{k_{l+1}}-I_{k_{l+1}z})%
\bigotimes ...\bigotimes (\frac{1}{2}E_{k_{n}}-I_{k_{n}z}).\ \ \qquad \ \
(32)$ \newline
Obviously, the operators $[X_{l},(D_{0}-D_{N-1})]_{\pm }$ are $l-$order
quantum operators. Therefore, if the integer index $l$ is an even number the
operators $[X_{l},(D_{0}-D_{N-1})]_{\pm }$ will be even-order
multiple-quantum operators. Note that when the integer index $l$ is an even
number the Hermitian operator $%
X_{l}=2^{l-1}I_{k_{1}x}I_{k_{2}x}...I_{k_{l}x} $ also is really an
even-order multiple-quantum operator which is a linear combination of the
even-order multiple-quantum operators with various quantum orders smaller
than $l$ including zero-quantum operators. In a more general case an $l-$%
order multiple-quantum operator $H_{l}$ could be further generated from the
Hermitian operators $i[X_{l},(D_{0}\pm D_{N-1})]$ and $[X_{l},(D_{0}\pm
D_{N-1})]_{+},$ for example,

$H_{l}=\stackrel{}{\stackunder{l}{\sum }}\alpha _{l}^{\pm }[iX_{l},(D_{0}\pm
D_{N-1})]+\stackrel{}{\stackunder{l}{\sum }}\beta _{l}^{\pm
}[X_{l},(D_{0}\pm D_{N-1})]_{+},\qquad \qquad \ \ $

$H_{l}=\exp (-i\pi I_{k_{l+j}x})[iX_{l},(D_{0}\pm D_{N-1})]\exp (i\pi
I_{k_{l+j}x})$ $(j=1,2,...,n-l).$\newline
To load the information of the marked state $|s\rangle $ on the
multiple-quantum operators one may first prepare the operators $\{X_{l}\}$
or the operator function $f(D_{s})$ from the diagonal operator $D_{s}$ of
the marked state $|s\rangle $ and other diagonal operators by Eq.(26) or
(27). Then these operators are further used to construct the desired
multiple-quantum operators in a similar way to $[X_{l},(D_{0}-D_{N-1})]_{\pm
}$. Once the desired Hermitian multiple-quantum operator $H_{l}$ is prepared
one can further construct the multiple-quantum unitary operator $%
U_{l}(D_{s},\lambda _{l})=\exp (-i\lambda _{l}H_{l})$. There is a well-known
operator identical relation for any two Hermitian operators $A$ and $B$ [33,
34]:

$\stackunder{m\rightarrow \infty }{\stackrel{}{\lim }}(e^{iA/\sqrt{m}}e^{iB/%
\sqrt{m}}e^{-iA/\sqrt{m}}e^{-iB/\sqrt{m}})^{m}=e^{-[A,B]}.\qquad \qquad
\qquad \qquad \quad (33)$\newline
This formula and Eq.(30) could be helpful for the decomposition of the
multiple-quantum unitary operator into a sequence of the oracle unitary
operations and oracle-independent unitary operations. This decomposition
could be used as the oracle iterative sequence $U(D_{s})$ in the quantum
search sequence.\newline
\newline
\textbf{3.3 The Grover iterative sequence in the spin ensemble}

According the above scheme of Eq.(24) to construct the oracle unitary
sequences $U(D_{s})$ a simple oracle unitary sequence $U(D_{s})$ is
constructed explicitly below. This sequence is similar to the Grover
iterative sequence [3, 17]. The time evolution propagator for the oracle
unitary sequence (here denote $U(D_{s})$ as $U(m)$) is constructed simply by

$\qquad \qquad U(m)=[\exp (-i\pi D_{N-1})\exp (-i\pi D_{s}^{x})]^{m}\qquad
\quad \ \qquad \qquad \quad \ \ (34)$ \newline
where the oracle-independent unitary operation $\exp (-i\pi D_{N-1})$ is a $%
LOMSO$ unitary operator and the oracle unitary operation sequence $\exp
(-i\pi D_{s}^{x})$ a multiple-quantum unitary operator. The Hermitian
multiple-quantum operator $D_{s}^{x}$ is defined as

$D_{s}^{x}=\exp (-i\frac{\pi }{2}F_{y})D_{s}\exp (i\frac{\pi }{2}F_{y})=%
\stackrel{n}{\stackunder{k=1}{\bigotimes }}(\frac{1}{2}%
E_{k}+a_{k}^{s}I_{kx}).\qquad \qquad \quad \ \qquad \ \ (35)$ \newline
By using the relation between the two diagonal operators $D_{s}$ and $D_{0}$
[17],

$D_{s}=\exp (-i\frac{\pi }{2}F_{x})U_{ox}(-\frac{\pi }{2})D_{0}U_{ox}(\frac{%
\pi }{2})\exp (i\frac{\pi }{2}F_{x})\quad \quad \qquad \qquad \quad \ \
\qquad \ (36)$ \newline
where the auxiliary oracle unitary operation $U_{ox}(\theta )=\stackrel{n}{%
\stackunder{k=1}{\prod }}\exp [-i\theta a_{k}^{s}I_{kx}],$ one can
re-express the propagator of Eq.(34) as

$U(m)=\exp (-i\frac{\pi }{2}F_{y})\exp (-i\frac{\pi }{2}F_{x})U_{ox}(-\frac{%
\pi }{2})$

$\times [\exp (-i\pi D_{0}^{x})\exp (-i\pi D_{0})]^{m}U_{ox}(\frac{\pi }{2}%
)\exp (i\frac{\pi }{2}F_{x})\exp (i\frac{\pi }{2}F_{y}).\quad \qquad \quad \
\ (37)$ \newline
It follows from the definition of the diagonal operator $D_{s}$ of the
marked state that there is an expansion for the unitary operator $\exp (\pm
i\pi D_{s})$ [17],

$\ \qquad \qquad \exp (\pm i\theta D_{s})=E+(-1+\exp (\pm i\theta
))D_{s}.\qquad \qquad \qquad \quad \ \qquad (38)$ \newline
The iterative part of the propagator $U(m)$ of Eq.(37) then can be
simplified by the expansion

$G(m)=[\exp (-i\pi D_{0}^{x})\exp (-i\pi D_{0})]^{m}$

$\ \ \qquad =[E-2D_{0}-2D_{0}^{x}+4D_{0}^{x}D_{0}]^{m}\qquad \qquad \qquad
\qquad \qquad \qquad \ \ \quad \ (39)$ \newline
It can turn out based on the definition of the diagonal operator $D_{k}$
that the operator set $\{E,D_{0},D_{0}^{x},D_{0}D_{0}^{x},D_{0}^{x}D_{0}\}$
form a closed operator algebra subspace. Then the iterative sequence of
Eq.(39) can be generally written as

$G(m)=E+\alpha _{1}(m)D_{0}+\alpha _{2}(m)D_{0}^{x}+\alpha
_{3}(m)D_{0}D_{0}^{x}+\alpha _{4}(m)D_{0}^{x}D_{0}.\qquad (40)$ \newline
Using the relations $G(m+1)=G(1)G(m)=G(m)G(1)$ one can obtain the recursion
equations for the coefficients $\{\alpha _{k}(m)\},$

$\left( 
\begin{array}{l}
\alpha _{1}(m+1) \\ 
\alpha _{3}(m+1)
\end{array}
\right) =\left( 
\begin{array}{ll}
-1+4/N & 2/N \\ 
-2 & -1
\end{array}
\newline
\right) \left( 
\begin{array}{l}
\alpha _{1}(m) \\ 
\alpha _{3}(m)
\end{array}
\right) +\left( 
\begin{array}{l}
-2 \\ 
0
\end{array}
\right) ,\quad \ \ (41a)$

$\left( 
\begin{array}{l}
\alpha _{2}(m+1) \\ 
\alpha _{4}(m+1)
\end{array}
\right) =\left( 
\begin{array}{ll}
-1 & -2/N \\ 
2 & -1+4/N
\end{array}
\newline
\right) \left( 
\begin{array}{l}
\alpha _{2}(m) \\ 
\alpha _{4}(m)
\end{array}
\right) +\left( 
\begin{array}{l}
-2 \\ 
4
\end{array}
\right) \qquad \ (41b)$ \newline
and $\alpha _{1}(m)=\alpha _{2}(m)$ and $\alpha _{4}(m)=-(2\alpha
_{1}(m)+\alpha _{3}(m)).$ The recursion equations (41a) and (41b) have the
following solution:

$\alpha _{1}(m)=\alpha _{2}(m)=-\QDABOVE{1pt}{N}{N-1}[1-\cos (m\theta )],$

$\alpha _{3}(m)=-\QDABOVE{1pt}{N}{N-1}[-1+\cos (m\theta )+\sqrt{N-1}\sin
(m\theta )],$

$\alpha _{4}(m)=\QDABOVE{1pt}{N}{N-1}[1-\cos (m\theta )+\sqrt{N-1}\sin
(m\theta )]$\newline
where $\cos \theta =-1+2/N$ and $\sin \theta =2\sqrt{N-1}/N.$

The time evolution of a spin ensemble under the propagator of Eq.(37) then
can be expressed as

$\rho _{I}(m)=U(m)\rho _{I}(0)U(m)^{-1}$

$=\exp (-i\frac{\pi }{2}F_{y})\exp (-i\frac{\pi }{2}F_{x})U_{ox}(-\frac{\pi 
}{2})G(m)$

$\times U_{ox}(\frac{\pi }{2})\exp (i\frac{\pi }{2}F_{x})\exp (i\frac{\pi }{2%
}F_{y})\rho _{I}(0)\exp (-i\frac{\pi }{2}F_{y})\exp (-i\frac{\pi }{2}%
F_{x})U_{ox}(-\frac{\pi }{2})$

$\times G(m)^{+}U_{ox}(\frac{\pi }{2})\exp (i\frac{\pi }{2}F_{x})\exp (i%
\frac{\pi }{2}F_{y}).\qquad \quad \qquad \ \ \qquad \qquad \qquad \qquad \
(42)$ \newline
Now assume that the initial density operator is given by $\rho _{I}(0)=%
\stackrel{n}{\stackunder{k}{\sum }}\varepsilon _{k}I_{kz}.$ Then one has the
unitary transformation relations:

$U_{ox}(\frac{\pi }{2})\exp (i\frac{\pi }{2}F_{x})\exp (i\frac{\pi }{2}%
F_{y})\rho _{I}(0)\exp (-i\frac{\pi }{2}F_{y})\exp (-i\frac{\pi }{2}%
F_{x})U_{ox}(-\frac{\pi }{2})$

$=-\stackrel{n}{\stackunder{k}{\sum }}\varepsilon _{k}I_{kx},$

$G(m)\{-\stackrel{n}{\stackunder{k}{\sum }}\varepsilon _{k}I_{kx}\}G(m)^{+}=-%
\stackrel{n}{\stackunder{k}{\sum }}\varepsilon _{k}\{I_{kx}+\gamma
_{1}(m)D_{0}+\gamma _{2}(m)D_{0}^{x}$

$+\gamma _{3}(m)D_{0}D_{0}^{x}+\gamma _{4}(m)D_{0}^{x}D_{0}+\gamma
_{5}(m)D_{0}I_{kx}+\gamma _{6}(m)I_{kx}D_{0}$

$+\gamma _{7}(m)D_{0}^{x}D_{0}I_{kx}+\gamma _{8}(m)I_{kx}D_{0}D_{0}^{x}\}$%
\newline
where the coefficients $\gamma _{l}(m)$ $(l=1,2,...,8)$ are given by

$\gamma _{1}(m)=\frac{1}{2N}[\alpha _{1}(m)^{*}\alpha _{3}(m)+\alpha
_{1}(m)\alpha _{3}(m)^{*}+\alpha _{3}(m)^{*}\alpha _{3}(m)],$

$\gamma _{2}(m)=\frac{1}{2}[\alpha _{2}(m)+\alpha _{2}(m)^{*}+\alpha
_{2}(m)^{*}\alpha _{2}(m)$

$+\frac{1}{N}\alpha _{2}(m)^{*}\alpha _{4}(m)+\frac{1}{N}\alpha
_{2}(m)\alpha _{4}(m)^{*}],$

$\gamma _{3}(m)=\frac{1}{2}[\alpha _{3}(m)+\alpha _{1}(m)\alpha
_{2}(m)^{*}+\alpha _{2}(m)^{*}\alpha _{3}(m)+\frac{1}{N}\alpha _{3}(m)\alpha
_{4}(m)^{*}],$

$\gamma _{4}(m)=\frac{1}{2}[\alpha _{3}(m)^{*}+\alpha _{1}(m)^{*}\alpha
_{2}(m)+\alpha _{2}(m)\alpha _{3}(m)^{*}+\frac{1}{N}\alpha _{3}(m)^{*}\alpha
_{4}(m)],$

$\gamma _{5}(m)=\alpha _{1}(m),\gamma _{6}(m)=\alpha _{1}(m)^{*},$

$\gamma _{7}(m)=\alpha _{4}(m),\gamma _{8}(m)=\alpha _{4}(m)^{*}.$\newline
The density operator $\rho _{I}(m)$ of Eq.(42) then can be further given by

$\rho _{I}(m)=\stackrel{n}{\stackunder{k=1}{\sum }}\varepsilon
_{k}\{I_{kz}-\gamma _{1}(m)D_{s}^{x}-\gamma _{2}(m)D_{N-1}$

$-\gamma _{3}(m)D_{s}^{x}D_{N-1}-\gamma _{4}(m)D_{N-1}D_{s}^{x}+\gamma
_{5}(m)D_{s}^{x}I_{kz}$

$+\gamma _{6}(m)I_{kz}D_{s}^{x}+\gamma _{7}(m)D_{N-1}D_{s}^{x}I_{kz}+\gamma
_{8}(m)I_{kz}D_{s}^{x}D_{N-1}\}.\qquad \quad \quad \quad \ \ (43)$ \newline
Now the conversion efficiency of the initial density operator $\rho _{I}(0)=%
\stackrel{n}{\stackunder{k}{\sum }}\varepsilon _{k}I_{kz}$ to the
multiple-quantum coherence under the Grover iterative sequence of Eq.(34)
can be obtained from Eq.(43). The $LOMSO$ operator in the density operator $%
\rho _{I}(m)$ of Eq.(43) which have not any contribution to the
multiple-quantum transitions is given by

$\rho _{z}(m)=\stackrel{n}{\stackunder{k=1}{\sum }}\varepsilon _{k}\{I_{kz}-%
\frac{1}{N}\gamma _{1}(m)E-\gamma _{2}(m)D_{N-1}-\frac{1}{N}\gamma
_{3}(m)D_{N-1}$\newline
$-\frac{1}{N}\gamma _{4}(m)D_{N-1}+\frac{1}{N}\gamma _{5}(m)I_{kz}+\frac{1}{N%
}\gamma _{6}(m)I_{kz}+\frac{1}{2N}\gamma _{7}(m)D_{N-1}+\frac{1}{2N}\gamma
_{8}(m)D_{N-1}\}.\qquad \qquad \qquad \quad \ $\newline
The density operator $\rho _{I}(m)$ is traceless since the initial density
operator $\rho _{I}(0)$ is traceless. Then the $LOMSO$ operator $\rho
_{z}(m) $ is reduced to the form

$\rho _{z}(m)=\stackrel{n}{\stackunder{k=1}{\sum }}\varepsilon
_{k}\{I_{kz}[1+\frac{1}{N}(\gamma _{5}(m)+\gamma _{6}(m))]-\frac{1}{N}\gamma
_{1}(m)E+\gamma _{1}(m)D_{N-1}\}.\quad (44)$\newline
In general, the diagonal operator $D_{s}$ defined by Eq.(9) can be expanded
as a sum of the $LOMSO$ product operator base

$D_{s}=\frac{1}{N}\{E+\stackrel{n}{\stackunder{k=1}{\sum }}%
(a_{k}^{s}2I_{kz})+\stackrel{n}{\stackunder{l>k=1}{\sum }}%
(a_{k}^{s}2I_{kz})(a_{l}^{s}2I_{lz})$

$+\stackrel{n}{\stackunder{m>l>k=1}{\sum }}%
(a_{k}^{s}2I_{kz})(a_{l}^{s}2I_{lz})(a_{m}^{s}2I_{mz})$

$+......+(a_{1}^{s}2I_{1z})(a_{2}^{s}2I_{2z}).....(a_{n}^{s}2I_{nz})\}.%
\qquad \qquad \qquad \qquad \qquad \qquad \qquad (45)$\newline
Then inserting the expansion (45) of the diagonal operator $D_{N-1}$ into
Eq.(44) one obtains

$\rho _{z}(m)=\stackrel{n}{\stackunder{k=1}{\sum }}\varepsilon _{k}I_{kz}[1+%
\frac{1}{N}(\gamma _{5}(m)+\gamma _{6}(m))]$

$+\frac{1}{N}(\stackrel{n}{\stackunder{k=1}{\sum }}\varepsilon _{k})\gamma
_{1}(m)\{\stackrel{n}{\stackunder{k=1}{\sum }}(-2I_{kz})+\stackrel{n}{%
\stackunder{l>k=1}{\sum }}(-2I_{kz})(-2I_{lz})$

$+\stackrel{n}{\stackunder{m>l>k=1}{\sum }}(-2I_{kz})(-2I_{lz})(-2I_{mz})$

$+......+(-2I_{1z})(-2I_{2z}).....(-2I_{nz})\}.\qquad \qquad \qquad \qquad
\qquad \qquad \qquad \ (46)$ \newline
Therefore, the conversion coefficient $C_{m}$ for the initial density
operator (or the initial longitudinal magnetization) $\rho _{I}(0)=\stackrel{%
n}{\stackunder{k}{\sum }}\varepsilon _{k}I_{kz}$ is given by

$C_{m}=1+\frac{1}{N}[\gamma _{5}(m)+\gamma _{6}(m)]-\frac{2}{N}(\stackrel{n}{%
\stackunder{l=1}{\sum }}\varepsilon _{l}/\varepsilon _{k})\gamma
_{1}(m).\qquad \qquad \qquad \ \qquad (47)$ \newline
The conversion coefficient $C_{m}=1$ means that the initial magnetization $%
\rho _{I}(0)$ is not transferred into any other components including
multiple-quantum coherence in the spin ensemble, while $C_{m}=0$ means that
the initial magnetization $\rho _{I}(0)$ is completely converted into other
components in the spin ensemble. It can be seen from the coefficients $%
\{\alpha _{k}(m)\}$ that the maximum $\gamma _{1}(m),$ $\gamma _{5}(m),$ $%
\gamma _{6}(m)\sim O(1)$ and occur at $m\sim \sqrt{N}$. Then the conversion
coefficient $C_{m}$ of Eq.(47) shows that the initial magnetization $\rho
_{I}(0)$ can not be transferred effectively into multiple-quantum coherence
for a large qubit number ($N=2^{n}$) and the transferred efficiency $%
(1-C_{m})$ decreases exponentially as the qubit number for any iterative
number $m$. This result shows that the Grover iterative sequence (34) is not
an effective sequence to enhance the NMR signal and amplify the realizable
size of a spin ensemble in solving quantum search problem in a spin
ensemble. This is quite different from the pure-state quantum search
algorithm which can achieve a quadratic amplification for the amplitude of
the marked quantum state. One sees again the difference between the ensemble
quantum search computation and the pure-state quantum search computation.%
\newline
\newline
\textbf{3.4 NMR signal enhancement by cross interaction between
oracle-dependent and oracle-independent interactions}

The extraction of the solution information of the search problem in the
multiple-quantum spectrum could be simplified by preparing the suitable
oracle iterative sequence $U(D_{s})$ and its effective Hamiltonian $H(D_{s})$
used to excite the multiple-quantum coherence.. As an example, assume that
the information of the marked state $|s\rangle $ is loaded on the
zero-quantum operator subspace of the Liouville operator space of the spin
system. The preparation for the zero-quantum Hamiltonian carrying the
information of the marked state can be carried out by using the discrete
Fourier analysis and the phase cycling technique [17], as can be seen in
Eq.(28). By starting the proper operator function $f(D_{s})$ of the diagonal
operator $D_{s}$ of the marked state and the oracle-independent operator
function $f(D_{r})$ the zero-quantum Hamiltonian may be constructed generally

$H_{zq}=\frac{1}{N_{1}}\stackrel{N_{1}-1}{\stackunder{k=0}{\sum }}\exp
(-i\varphi _{k}F_{z})[f(D_{s})+f(D_{r})]\exp (-i\varphi _{k}F_{z})$

$\qquad =H_{zq}(D_{s})+H_{zq}(D_{r})\qquad \qquad \qquad \qquad \qquad
\qquad \qquad \qquad \qquad \ \ (48)$ \newline
and the corresponding zero-quantum unitary operator is already given by
Eq.(29). If the oracle-independent Hamiltonian $H_{zq}(D_{r})$ is suitably
chosen, for example, it may be chosen as those zero-quantum coherence of the 
$k-$qubit $(k\leq n)$ subsystem of the spin system and moreover, as the
dominating term in Eq.(48) with respect to the Hamiltonian $H_{zq}(D_{s}),$
then extraction of the solution information could be simplified and the
zero-quantum coherence which carry the information could be enhanced in
amplitude, as can be seen below. Now there are three types of the
zero-quantum coherence in the spin ensemble. One type is of the $k-$qubit
subsystem of the spin system, another of the rest $(n-k)-$qubit subsystem,
and the third is those cross zero-quantum coherence between the two
subsystem. The last type is more important. Assume that all spins of the $k-$%
qubit subsystem have the same precession frequency and so do all spins of
the rest $(n-k)-$qubit subsystem, but the two precession frequencies are
different each other. Then the cross zero-quantum coherence between the two
subsystems may have precession frequencies different from zero frequency
that is the precession frequency of the zero-quantum coherence of each
subsystem. Therefore, the cross zero-quantum coherence with nonzero
precession frequencies and those zero-quantum coherence of the two
subsystems can be discriminated in the zero-quantum spectrum. Since these
cross zero-quantum peaks also contain the information of the marked state
one could obtain the information of the marked state from the spectral
parameters of the cross zero-quantum peaks. The zero-quantum peaks of the $%
k- $qubit subsystem are strongest since the oracle-independent Hamiltonian $%
H_{zq}(D_{r})$ is the dominating term in the total Hamiltonian $H_{zq},$
while those peaks of the $(n-k)-$qubit subsystem are usually weakest and
even can not be observable if a small call number of the oracle unitary
operation is used in the sequence, but the information of the marked state
may not be obtained from these zero-quantum peaks because of overlapping of
the strong zero-quantum peaks without the information. However, amplitude
for the cross zero-quantum coherence is dependent on both the
oracle-dependent and oracle-independent Hamiltonians. Then the cross
zero-quantum peaks could be much stronger than those zero-quantum peaks of
the $(n-k)-$qubit subsystem and could be observable even when the peaks of
the $(n-k)-$qubit subsystem are too weak to be observable. Therefore, one
may use these cross zero-quantum peaks with nonzero precession frequencies
to determine the solution of the search problem. Obviously, this NMR signal
enhancement is achieved by the cross interaction between the
oracle-dependent and oracle-independent Hamiltonians. To see more clearly
the enhancement mechanism one first expresses the zero-quantum unitary
transformation in the interaction frame as

$U_{zq}(t)=\exp (-iH_{zq}t)=\exp [-itH_{zq}(D_{r})]T\exp (-i\stackrel{t}{%
\stackunder{0}{\int }}H_{zq}^{^{\prime }}(D_{s},t^{\prime })dt^{\prime
})\quad (49)$ \newline
where $T$ is Dyson time-ordering operator and the oracle-dependent
Hamiltonian in the interaction frame is given by

$\qquad H_{zq}^{^{\prime }}(D_{s},t)=\exp [itH_{zq}(D_{r})]H_{zq}(D_{s})\exp
[-itH_{zq}(D_{r})].\ \qquad \qquad \ \ (50)$ \newline
Equation (49) may be used to build up the quantum circuit $U_{zq}(t)$ which
equals $U(D_{s})$ in the quantum search sequence [37]. The interaction
Hamiltonian (50) is responsible for generating the cross zero-quantum peaks.
It can be expanded formally as

$H_{zq}^{^{\prime }}(D_{s},t)=H_{zq}(D_{s})+it[H_{zq}(D_{r}),$ $%
H_{zq}(D_{s})]$

$\qquad \qquad \quad \quad -\frac{1}{2}t^{2}[H_{zq}(D_{r}),$ $%
[H_{zq}(D_{r}), $ $H_{zq}(D_{s})]]+....\qquad \qquad \qquad \ \ (51)$\newline
One sees that there are the cross interaction terms between the two
Hamiltonians $H_{zq}(D_{r})$ and $H_{zq}(D_{s})$ in the higher-order terms
of the expansion except the term $H_{zq}(D_{s})$. These cross interaction
terms are responsible for generating the cross zero-quantum peaks.
Obviously, the cross interaction in strength is dependent on the oracle
independent Hamiltonian $H_{zq}(D_{r}),$ and a strong Hamiltonian $%
H_{zq}(D_{r})$ will generate a strong cross interaction. Therefore, a strong
oracle-independent Hamiltonian would enhance greatly the cross zero-quantum
peaks since strong cross interaction would create strong cross zero-quantum
peaks. On the other hand, the oracle-independent Hamiltonian $H_{zq}(D_{r})$
becomes more and more important in the higher order terms, as can be seen in
Eq.(51). A cross interaction term with a high order such as $%
[H_{zq}(D_{r}),...,$ $[H_{zq}(D_{r}),$ $H_{zq}(D_{s})]...]$ could amplify
the effect of the oracle- independent Hamiltonian on the cross zero-quantum
peaks.

Generally, the Baker-Campbell-Hausdorff (BCH) formula [38] can be exploited
to construct the higher-order cross interaction. For convenience, operators $%
A=H(D_{r})$ and $B=H(D_{s})$ denote the oracle-independent and
oracle-dependent Hamiltonians, respectively. The operator $A$ may be any
Hermitian operator of the spin system and usually is taken conveniently as
the $LOMSO$ operator. The oracle-dependent Hamiltonian $H(D_{s})$ is usually
taken as an Hermitian multiple-quantum operator. The BCH$\;$formula [35, 36,
38] shows that the symmetric composition for the two operators $A$ and $B$
with the second order approximation is written as

$S_{A}(x)=\exp (xA^{\prime })\exp (xB)\exp (xA^{\prime })=\exp
(xA_{1})\qquad \qquad \qquad \qquad \ (52a)$ \newline
with the Hermitian operator,

$xA_{1}=x(A+B)+\frac{1}{6}x^{3}([B,[B,A^{\prime }]]-[A^{\prime },[A^{\prime
},B]])+...,\qquad \qquad \qquad (52b)$\newline
where $x=it$ is the imaging time and $A^{\prime }=A/2$ (the same denoting
for operator $B^{\prime }$, $A_{1}^{\prime }$, ..., below)$.$ There is also
another second order symmetric composition with a different product order of
the operators $A$ and $B$,

$S_{B}(x)=\exp (xB^{\prime })\exp (xA)\exp (xB^{\prime })=\exp
(xB_{1})\qquad \qquad \qquad \ \qquad (53a)$ \newline
with the Hermitian operator,

$xB_{1}=x(A+B)+\frac{1}{6}x^{3}([A,[A,B^{\prime }]]-[B^{\prime },[B^{\prime
},A]])+....\qquad \qquad \ \ \quad (53b)$ \newline
Because the unitary operators $S_{q}(x)$ ($q=A,B$) of Eqs.(52a) and (53a)
satisfy the time symmetry: $S_{q}(x)S_{q}(-x)=E$ (E is unity operator) there
are not even-order commutators such as $[A,B]$, $[A,[A,[A,B]]]$, etc., in
the composed operators $A_{1}$ and $B_{1}$ of Eqs.(52b) and (53b) [36]. Here
an $n-$order cross interaction term is defined as a $(n+1)-$order commutator
of the operators $A$ and $B$. Then the second-order cross interaction, which
is defined as the third order commutators such as $[A,[A,B]]$ and $%
[B,[B,A]], $ can be built up in the same symmetric form as Eq.(52a) with the
operators $A_{1}$ and $B_{1}$ again according to the BCH formula,

$\exp (xA_{1}^{\prime })\exp (-xB_{1})\exp (xA_{1}^{\prime })=\exp
(xA_{3})\qquad \qquad \qquad \qquad \quad \qquad (54a)$\newline
where the cross interaction $A_{3}$ is given by

$xA_{3}=x(A_{1}-B_{1})+\frac{1}{6}x^{3}([-B_{1},[-B_{1},A_{1}^{\prime
}]]-[A_{1}^{\prime },[A_{1}^{\prime },-B_{1}]])+...$

$\qquad =\frac{1}{8}x^{3}([B,[B,A]]-[A,[A,B]])+...,\qquad \qquad \qquad
\qquad \qquad \qquad (54b)$\newline
and

$\exp (xB_{1}^{\prime })\exp (-xA_{1})\exp (xB_{1}^{\prime })=\exp
(xB_{3})\qquad \qquad \qquad \qquad \qquad \ \ \ (55a)$ \newline
where the cross interaction $B_{3}$ is given by

$xB_{3}=x(-A_{1}+B_{1})+\frac{1}{6}x^{3}([-A_{1},[-A_{1},B_{1}^{\prime
}]]-[B_{1}^{\prime },[B_{1}^{\prime },-A_{1}]])+...$

$\qquad =\frac{1}{8}x^{3}([A,[A,B]]-[B,[B,A]])+....\qquad \qquad \qquad
\qquad \qquad \qquad (55b)$\newline
It can be seen in Eqs.(54b) and (55b) that the operators $A$ and $B$
disappear and the dominating terms become the third order commutators $%
[A,[A,B]]$ and $[B,[B,A]]$ in the cross interaction $A_{3}$ and $B_{3}$. If
those commutators with order higher than four can be neglected in Eqs.(54b)
and (55b) then the operators $A_{3}$ and $B_{3}$ are approximated as the
second order cross interaction. The fourth-order cross interaction can be
prepared by starting the second order cross interaction $A_{3}$ and $B_{3}$,

$\qquad \qquad \exp (xA_{3}^{\prime })\exp (xB_{3})\exp (xA_{3}^{\prime
})=\exp (xA_{5}),\qquad \qquad \qquad \ \quad (56a)$ \newline
and

$\qquad \qquad \exp (xB_{3}^{\prime })\exp (xA_{3})\exp (xB_{3}^{\prime
})=\exp (xB_{5}).\qquad \qquad \qquad \quad \ (56b)$ \newline
The higher order cross interaction also can be easily built up with the same
method above. The complexity of the quantum circuit to construct the higher
order cross interaction is usually dependent on the two Hamiltonians $%
H(D_{r})$ and $H(D_{s}).$ However, it is easy to see that preparing $2m-$%
order cross interaction terms $A_{2m+1}$ and $B_{2m+1}$ need $\frac{1}{2}%
(3^{m+1}-1)$ and $\frac{1}{2}(3^{m+1}+1)$ oracle unitary sequences $\exp
(xB),$ respectively. How high order cross interaction $A_{2m+1}$ $(B_{2m+1})$
is required for an $n-$qubit spin system so that the initial density
operator $\rho _{I}(0)=\stackrel{n}{\stackunder{k}{\sum }}\varepsilon
_{k}I_{kz}$ can be transferred efficiently into the multiple-quantum
coherence by the oracle unitary sequence $U(D_{s})=\exp (-itA_{2m+1})$ ($%
\exp (-itB_{2m+1})$)? This is also dependent on the two Hamiltonians $%
H(D_{r})$ and $H(D_{s})$ but it could be lower than $n$ for the proper $%
H(D_{r})$ and $H(D_{s}).$ A much more efficient method to prepare the higher
order cross interaction can be seen in [35, 36]. According to the method
[35, 36] one first composes in a symmetric form the $(2m-1)-$order composed
unitary operators $f_{2m-1}^{A}(x)$ and $f_{2m-1}^{B}(x)$ by using the
second order composed unitary operators $S_{A}(x)$ and $S_{B}(x)$ of
Eq.(52a) and (53a), respectively,

$f_{2m-1}^{A}(x)=\exp [x(A+B)+x^{2m+1}A_{2m+1}+...]$

$\qquad \qquad =S_{A}(p_{1}x)S_{A}(p_{2}x)...S_{A}(p_{r}x)$,\quad \qquad $%
\qquad \qquad \qquad \qquad \qquad (57a)$

$f_{2m-1}^{B}(x)=\exp [x(A+B)+x^{2m+1}B_{2m+1}+...]$

$\qquad \qquad =S_{B}(p_{1}x)S_{B}(p_{2}x)...S_{B}(p_{r}x)\qquad \quad
\qquad \qquad \qquad \qquad \qquad \ (57b)$ \newline
where the parameters $\{p_{r}\}$ must be determined suitably and satisfy $%
\stackunder{k=1}{\stackrel{r}{\sum }}p_{k}=1,$ $p_{j}=p_{r+1-j}$ ($%
j=1,2,...,r-1$), and it is usually not easy to determine the parameters $%
\{p_{r}\}$ for a higher-order composition [35, 36]. In Eqs.(57a) and (57b)
it needs about $r\thickapprox 2^{m}$ unitary operators $S_{A}(x)$ ($S_{B}(x)$%
) to compose the unitary operator $f_{2m-1}^{A}(x)$ ($f_{2m-1}^{B}(x)$) for
small $m$ [35]. Note that there is difference of the product order of the
operators $A$ and $B$ in the operators $f_{2m-1}^{A}(x)$ and $%
f_{2m-1}^{B}(x).$ The difference is necessary to construct further the $2m-$%
order cross interaction with the operators $f_{2m-1}^{A}(x)$ and $%
f_{2m-1}^{B}(x).$ Again one uses the above symmetric BCH formula to compose
further the $2m-$order cross interaction:

$\exp (\frac{1}{2}x(A+B)+\frac{1}{2}x^{2m+1}B_{2m+1}+...)\exp
(-x(A+B)-x^{2m+1}A_{2m+1}+...)$

$\times \exp (\frac{1}{2}x(A+B)+\frac{1}{2}x^{2m+1}B_{2m+1}+...)$

$=\exp [-x^{2m+1}(A_{2m+1}-B_{2m+1})+...].\qquad \qquad \qquad \qquad \qquad
\qquad \quad \ \ (58)$ \newline
Obviously, here the parameter $x$ does not take a small value so that the $%
2m-$order cross interaction $A_{2m+1}$ or $B_{2m+1}$ could be large enough
to derive the conversion of the initial density operator to the
multiple-quantum coherence in a spin ensemble. This is different from the
conventional decomposition of exponential operators [33, 35, 36] where the
parameter $x$ is always taken as a small quantity.

Here leave two questions of the quantum search NMR multiple-quantum pulse
sequence in the future. One is how high order cross interaction is necessary
to excite effectively the multiple-quantum coherence in an $n-$qubit spin
ensemble. Another is how to extract experimentally the information of the
marked state from the multiple-quantum spectrum excited by the higher order
cross interaction. \newline
\[
\]
\newline
{\Large 4. Discussion}

The quantum mechanical unitary dynamical description for the Turing machines
was first given by Benioff [39] in 1980. This quantum Turing machine model
has not proved to be superior to the classical counterparts, as pointed out
by Deutsch [2]. However, the quantum mechanical unitary dynamics plays an
important role in computational speedup in ensemble quantum computation in
contrast to in the pure-state quantum computation. A direct result of the
quantum mechanical unitary dynamics is that the quantum search problem and
the parity problem could be solved in polynomial time on an NMR ensemble
quantum computer, while these problems are NP problems and can not be solved
efficiently in classical computation. Therefore these works really move the
frontier between the solvability and intractability of complex problems.
There is difference between ensemble quantum computation based on the
quantum mechanical unitary dynamics and the pure state version. For example,
in the paper it has been shown that the Grover algorithm is an ineffective
iterative sequence used to amplify the output signal of the NMR ensemble
quantum computation, while it can achieve a quadratic amplification for the
amplitude of the marked state in the pure-state quantum search. The reason
for this is that the iterative sequence of the Grover algorithm can amplify
quadratically the relative amplitude of the marked state and suppress any
other states in the quantum system, but the detectable NMR signal intensity
in a spin ensemble is not only proportional to the relative amplitude of the
marked state but also the ensemble-averaging magnetization strength of the
spin system. The ensemble quantum computation based on the quantum
mechanical unitary dynamics allows the separation each other between the
initial input state and the quantum circuit in a quantum algorithm. This
makes it possible for the highly mixed states of the spin ensemble to be
used in ensemble quantum computation. This also shows that there is
difference between new ensemble quantum computation and the classical NMR
quantum computation working on the concept of the pseudopure state or the
effective pure state.

In the paper it has been shown that the quantum search problem could be
solved in polynomial time on an NMR quantum computer. On the basis of the
quantum mechanical unitary dynamics the NMR multiple-quantum spectroscopy
has been developed to solve experimentally the quantum search problem. In
the quantum search sequence based on the NMR multiple-quantum spectroscopy
the information of the marked state is first loaded on the unitary evolution
propagator that excites the multiple-quantum coherence from the initial
density operator, i.e., a highly mixed state, of the spin ensemble. This
propagator is usually constructed with the oracle unitary operation and
oracle-independent unitary operations. The NMR multiple-quantum spectra that
carry the information of the marked state then are obtained experimentally.
Finally one could extract the information of the marked state from the
multiple-quantum spectra by the conventional NMR data processing. Although
number of the multiple-quantum coherence or transitions increases
exponentially as the qubit number in a spin ensemble, number of the
multiple-quantum spectral peaks can be controlled and can be as low as $%
(2n+1)$ in the multiple-quantum spectra for an $n-$qubit spin system$.$ This
ensures that at least some lower-order multiple-quantum peaks do not reduce
exponentially as the qubit number as long as the initial density operator is
transferred effectively to the multiple-quantum coherence by the oracle
iterative sequence of the quantum search sequence. The complexity of the NMR
multiple-quantum experiments is therefore mainly dependent on the oracle
iterative sequence exciting the multiple-quantum coherence and the
extraction of the information of the marked state from the multiple-quantum
spectra. The NMR\ multiple-quantum spectroscopy could provide a more
convenient and larger space to design new ensemble quantum computation
sequences or experiments. It also could provide possible approaches to
enhancing the output NMR signal of ensemble quantum computation and
extending the realizable size of the NMR ensemble quantum computation. In
particular, in the frame of the multiple-quantum spectroscopy the cross
interaction between oracle-dependent and oracle-independent interactions has
been proposed to enhance the output NMR signal of the ensemble quantum
computation. In comparison with those NMR quantum search algorithms or
experiments [11, 17, 40] the quantum search sequence based on the quantum
mechanical unitary dynamics and the NMR multiple-quantum spectroscopy would
be more powerful. 
\[
\]
\newline
{\Large References}\newline
1. C.H.Bennett and D.P.DiVincenzo, Nature 404, 247 (2000)\newline
2. D.Deutsch, Quantum theory, the Church-Turing principle and the universal
quantum computer, Proc.Roy.Soc.Lond. A 400, 97 (1985) \newline
3. L.K.Grover, Quantum mechanics helps in searching for a needle in a
haystack, Phys.Rev.Lett. 79, 325 (1997) \newline
4. E.Farhi and S.Gutmann, An analog analogue of a digital quantum
computation, Phys.Rev. A, 57, 2403 (1998) \newline
5. M.Boyer, G.Brassard, P.Hoyer, and A.Tapp, Tight bounds on quantum
searching, http://arXiv.gov/abs/quant-ph/9605034 (1996)\newline
6. C.H.Bennett, E.Bernstein, G.Brassard, and U.Vazirani, Strengths and
weaknesses of quantum computing, SIAM J.Computing.. 26, 1510 (1997)\newline
7. R.R.Ernst, G.Bodenhausen, and A.Wokaun, Principles of Nuclear Magnetic
Resonance in One and Two Dimensions (Oxford University Press, Oxford, 1987)%
\newline
8. R.Freeman, Spin Choreography, (Spektrum, Oxford, 1997)\newline
9. D.G.Cory, A.F.Fahmy, T.F.Havel, Proc.Natl.Acad.Sci. USA 94, 1634 (1997)%
\newline
10. N.A.Gershenfeld and I.L.Chuang, Science 275, 350 (1997) \newline
11. I.L.Chuang, N.A.Gershenfeld, and M.Kubinec, Phys.Rev.Lett. 80, 3408
(1998); J.A.Jones, M.Mosca, R.H.Hansen, Nature 393, 344 (1998)\newline
12. Z.L.Madi, R.Bruschweiler, and R.R.Ernst, One- and two-dimensional
ensemble quantum computing in spin Liouville space, J.Chem.Phys. 109, 10603
(1998) \newline
13. N.Linden, H.Barjat, and R.Freeman, Chem.Phys.Lett. 296, 61 (1998)\newline
14. M.A.Nielsen, E.knill, and R.Laflamme, Nature 396, 52 (1998)\newline
15. W.S.Warren, Science 277, 1688-1689 (1997) \newline
16. S.L.Braunstein, C.M.Caves, R.Jozsa, N.Linden, S.Popescu, and R.Schack,
Phys.Rev.Lett. 83, 1054 (1999) \newline
17. X.Miao, Universal construction for the unsorted quantum search
algorithms, http://arXiv.org/abs/quant-ph/0101126 (2001)\newline
18. X.Miao, A polynomial-time solution to the parity problem on an NMR
quantum computer, http://arXiv.org/abs/quant-ph/0108116 (2001)\newline
19. E.Farhi, J.Goldstone, S.Gutmann, and M.Sipser, A limit on the speed of
quantum computation in determining parity, Phys.Rev.Lett. 81, 5442 (1998)%
\newline
20. R.Beals, H.Buhrman, R.Cleve, M.Mosca, and R. de Wolf,

http://arXiv.org/abs/quant-ph/9802049 (1998)\newline
21. N.Gershenfeld and I.L.Chuang, Science 277, 1689-1690 (1997)\newline
22. L.J.Schulman and U.V.Vazirani, in Proceedings of the 31st Annual ACM
Symposium on Theory of Computing, 1999, pp.322\newline
23. P.O.Boykin, T.Mor, V.Roychowdhury, F.Vatan, and R.Vrijen,

http://xxx.lanl.gov/abs/quant-ph/0106093\newline
24. X.Miao, Multiple-quantum operator algebra spaces and description for the
unitary time evolution of multilevel spin systems, Molec.Phys. 98, 625
(2000) \newline
25. X.Miao, Universal construction of unitary transformation of quantum
computation with one- and two-body interactions, \newline
http://arXiv.gov/abs/quant-ph/0003068 (2000) \newline
26. R.Cleve, A.Ekert, C.Macchiavello, and M.Mosca, Proc.Roy.Soc.Lond. A 454,
339 (1998)\newline
27. D.Beckman, A.N.Chari, S.Devabhaktuni, and J.Preskill, Efficient networks
for quantum factoring, Phys.Rev. A 54, 1034 (1996) \newline
28. X.Miao, A convenient method to prepare the effective pure state in a
quantum ensemble, http://arXiv.gov/abs/quant-ph/0008094 (2000) \newline
29. A.L.Davis, G.Estcourt, J.Keeler, E.D.Laue, and J.J.Titman,

J.Magn.Reson. A105, 167 (1993)\newline
30. Y.S.Yen and A.Pines, Multiple-quantum NMR in solids, J.Chem.Phys. 78,
3579 (1983) \newline
31. W.S.Warren, D.P.Weitekamp, and A.Pines, Theory of selective excitation
multiple-quantum transitions, J.Chem.Phys. 73, 2084 (1980) and therein 
\newline
32. H.F.Trotter, On the product of semigroups of operators,

Proc.Am.Math.Soc. 10, 545 (1959)\newline
33. M.Suzuki, Decomposition formulas of exponential operators and Lie
exponentials with some applications to quantum mechanics and statistical
physics, J.Math.Phys. 26, 601 (1985) \newline
34. R.M.Wilcox, J.Math.Phys. 8, 108 (1967)\newline
35. M.Suzuki, Fractal decomposition of exponential operators with
applications to many-body theories and Monte Carlo simulations, Phys.Lett.
A, 146, 319 (1990); A 165, 387 (1992)\newline
36. H.Yoshida, Construction of higher order symplectic integrators,
Phys.Lett. A 150, 262 (1990)\newline
37. S.Wiesner, Simulation of many-body quantum systems by a quantum
computer, http://arXiv.org/abs/quant-ph/9603028 (1996) \newline
38. Z.D.Yan and Y.C.Xu, Lie Group and Lie Algebra, (the Higher Education
Press, Beijing), 1985 (in Chinese)\newline
39. P.Benioff, The computer as a physical system: A microscopic quantum
mechanical Hamiltonian model of computers as represented by Turing machines,
J.Statist.Phys. 22, 563 (1980); Quantum mechanical Hamiltonian models of
Turing machines, J.Statist.Phys. 29, 515 (1982)\newline
40. R.Bruschweiler, Novel strategy for Database searching in spin Liouville
space by NMR ensemble computing, Phys. Rev. Lett. 85, 4815 (2000) \newline
\newline

\end{document}